\documentclass[aps,pra,reprint,twocolumn,superscriptaddress]{revtex4-1}

\usepackage[T1]{fontenc}
\usepackage{lmodern}
\usepackage[utf8]{inputenc}
\usepackage[english]{babel}

\usepackage{microtype}
\usepackage{indentfirst}	

\usepackage{amsthm, amssymb, amstext, amsmath, bbm}
\usepackage{dsfont}

\usepackage{float}
\usepackage{tikz}

\usepackage{epstopdf}
\usepackage{blkarray} 
\usepackage{MnSymbol}

\usepackage[normalem]{ulem}

\usepackage{varioref}
\usepackage[colorlinks=true,
							pdfstartview=FitV, 
							linkcolor=blue,  
							citecolor=blue, 
							urlcolor=blue,
							linktocpage=true]{hyperref}
\usepackage{braket}

\newcommand{\be}{\begin{equation}}
\newcommand{\ee}{\end{equation}}
\newcommand{\ba}{\begin{eqnarray}}
\newcommand{\ea}{\end{eqnarray}}

\newcommand{\ii}{{\rm i}}
\newcommand{\Tr}[1]{\mathrm{Tr}\!\left[#1\right]}

\newcommand{\supmat}[1]{\bar{\bar{#1}}}

\newcommand{\sss}{\hat{\rho}_{ss}}
\newcommand{\eig}[1]{\hat{\rho}_{#1}}
\renewcommand{\L}{\mathcal{L}}
\newcommand{\Lmat}{\bar{\bar{\mathcal{L}}}}
\newcommand{\symmat}[1]{\hat{\tilde{\rho}}_{#1}}

\makeatletter
\newcommand*\bigcdot{\mathpalette\bigcdot@{.5}}
\newcommand*\bigcdot@[2]{\mathbin{\vcenter{\hbox{\scalebox{#2}{$\m@th#1\bullet$}}}}}
\makeatother

\renewcommand{\Re}[1]{\mathrm{Re}\left[#1\right]}
\renewcommand{\Im}[1]{\mathrm{Im}\left[#1\right]}

\newcommand{\abs}[1]{\lvert #1 \rvert}

\begin{document}

\title{Spectral theory of Liouvillians for dissipative phase transitions}

\author{Fabrizio Minganti}
\affiliation{Laboratoire Mat\'{e}riaux et Ph\'{e}nom\`{e}nes Quantiques, Universit\'{e} Paris Diderot, CNRS-UMR7162, 75013 Paris, France}

\author{Alberto Biella}
\affiliation{Laboratoire Mat\'{e}riaux et Ph\'{e}nom\`{e}nes Quantiques, Universit\'{e} Paris Diderot, CNRS-UMR7162, 75013 Paris, France}

\author{Nicola Bartolo}
\affiliation{Laboratoire Mat\'{e}riaux et Ph\'{e}nom\`{e}nes Quantiques, Universit\'{e} Paris Diderot, CNRS-UMR7162, 75013 Paris, France}

\author{Cristiano Ciuti}
\email{cristiano.ciuti@univ-paris-diderot.fr}
\affiliation{Laboratoire Mat\'{e}riaux et Ph\'{e}nom\`{e}nes Quantiques, Universit\'{e} Paris Diderot, CNRS-UMR7162, 75013 Paris, France}
\date{\today}
 
 \begin{abstract}
A state of an open quantum system is described by a density matrix, whose dynamics is governed by a Liouvillian superoperator.
Within a general framework, we explore fundamental properties of both first-order dissipative phase transitions and second-order dissipative phase transitions associated with a symmetry breaking.
In the critical region, we determine the general form of the steady-state density matrix and of the Liouvillian eigenmatrix whose eigenvalue defines the Liouvillian spectral gap.
We illustrate our exact results by studying some paradigmatic quantum optical models exhibiting critical behavior.
 \end{abstract}

\maketitle



\section{Introduction and motivations}
In classical physics, {phase} transitions are driven by a competition between the value of the system energy and the entropy produced by its thermal fluctuations \cite{LandauBOOK69}.
A quantum system at zero temperature has zero entropy and is in its ground state, which is the state minimizing the system energy \cite{SachdevBOOK01}. 
However, critical phenomena can occur in the thermodynamic limit as a result of the competition between non commuting terms of the Hamiltonian. 

Driven-dissipative systems have an intrinsic non-equilibrium nature and the properties of the stationary state of the system can not be determined via a free energy analysis \cite{CarmichaelBOOK98,GardinerBOOK04,HarocheBOOK06,Weiss_book,Daley}.
The statistical mechanics of such systems can be remarkably rich.
For example, classical systems \cite{Zia_1995} can display long-range order in 2D \cite{Toner_1988}, since their driven-diffusive nature can violate the Mermin-Wagner theorem \cite{Mermin_1966}, which is valid at equilibrium. 
At a quantum level, by properly designing the coupling with the environment and the driving mechanisms, it is possible to stabilize phases without an equilibrium counterpart \cite{LeePRL13,JinPRX16}. 

The impressive experimental advances of the last decade provide the  opportunity to explore non-equilibrium critical phenomena on a variety of platforms. 
Lattices of superconducting resonators \cite{HouckNatPhys12,FitzpatrickPRX17}, Rydberg atoms in optical lattices \cite{Mueller_2012,BernienNAT2017}, optomechanical systems \cite{MarkusRev2014,GilSantosPRL17}, and exciton-polariton condensates \cite{KasprzakNAT2006,CarusottoRMP13} provide a highly-controllable playground in which to study the emergence of dissipative phase transitions. 
In the thermodynamic limit, the competition between Hamiltonian evolution, pumping and dissipation processes can trigger a non-analytical change in the steady state \cite{KesslerPRA12}.
The engineering of complex many-body phases has been deeply explored in different contexts \cite{DiehlNATPH2008,VerstraeteNATPH2009}. Dissipative phase transitions have been discussed theoretically for photonic systems \cite{CarmichaelPRX15,WeimerPRL2015,BenitoPRA16,MendozaPRA16,CasteelsPRA16,BartoloPRA16,CasteelsPRA17,CasteelsPRA17-2,Foss-FeigPRA17,BiondiPRA2017,BiellaPRA2017,SavonaPRA2017,Munuz2018}, lossy polariton condensates \cite{SiebererPRL13,SiebererPRB14,AltmanPRX15}, and spin models \cite{LeePRA2011,KesslerPRA12,LeePRL13,ChanPRA2015,JinPRX16,MaghrebiPRB16,RotaPRB17,OverbeckPRA17,RoscherarXiv18}.

The interplay between classical and quantum fluctuations in triggering nonequilibrium phase transitions has been addressed by different methods, including renormalization group approaches based on the Keldysh formalism \cite{TorrePRB2012, SiebererPRL13,MarinoPRL2016} and via extensive numerical analysis of lattice systems \cite{RotaPRB17,VicentiniPRA2018,RotaNJP18}.
Very recently, the critical properties have been investigated also experimentally in single superconducting cavities \cite{FinkPRX17}, semiconductor micropillars \cite{RodriguezPRL17,FinkNATPHY18}, and large arrays of microwave cavities \cite{FitzpatrickPRX17}.
Our understanding of criticality in such systems, however, is still in its infancy, since their description cannot be traced back to the traditional framework of equilibrium statistical mechanics.

One of the first investigations in this direction was reported by Kessler {\it et al.}~\cite{KesslerPRA12}.
They considered a specific spin model, conjecturing some general properties of dissipative phase transitions.
However, a general theory connecting Liouvillian spectral properties and dissipative phase transitions is still lacking.
In this work, we wish to provide a common theoretical framework to describe the emergence of critical behavior in Markovian open quantum systems, analyzing both first- and second-order phase transitions. 
We show the general form of the steady-state density matrix in the vicinity of the critical point. We determine also the form of the eigenmatrix of the Liouvillian superoperator corresponding to the non-zero eigenvalue with the smallest modulus of the real part (the so-called Liouvillian spectral gap or asymptotic decay rate).
When the transition is of the first order, we show that the gap closes only at the critical point, where the stationary state is bimodal.
Concerning second-order phase transitions associated with a symmetry-breaking, we provide a general spectral description proving that the Liouvillian gap remains closed in the whole region of broken symmetry.
In this context, we highlight the connection between the structure of the eigenmatrices and the symmetry properties of the Lindblad master equation.
Note that, according to the theory presented in this work, when the Liouvillian gap closes, also the imaginary part of the corresponding eigenvalue must vanish.
This is a more stringent constraint with respect to that discussed in \cite{KesslerPRA12}, where only the real part is assumed to vanish.
Particular attention is devoted to the connection between our results and their relation to mean-field solutions, bistability phenomena and metastability.
We bring under a common paradigm apparently different phenomena related to dissipative phase transitions which have been observed experimentally \cite{FitzpatrickPRX17, FinkPRX17,FinkNATPHY18} and predicted theoretically \cite{GutierrezArxiv18,JinPRX16,RotaNJP18,VicentiniPRA2018} for specific models.
We remark that our results are model-independent.
One of the goals of the present work is to identify a general spectral mechanism which can explain these phenomena regardless of the nature of the system (bosons, fermions, or spins) and dimensionality.

In order to better illustrate our general results, we analyze some specific paradigmatic cases of linearly- \cite{DrummondJPA80} and quadratically-driven \cite{MingantiSciRep16,BartoloPRA16,BartoloEPJST17} Kerr resonators.
Those models are known to undergo a phase transition (of first- \cite{CasteelsPRA17-2} and second-order \cite{BartoloPRA16}, respectively) in the thermodynamic limit of large excitation numbers.

This paper is structured as follows.  
In Sec.~\ref{subsec:proprieties} we introduce the theoretical framework, pointing out some general key properties of the Liouvillian superoperator and of density matrices. 
In Sec.~\ref{Sec:First_order_Phase_transition} and Sec.~\ref{Sec:Secon_order} we consider, respectively, first- and second-order dissipative phase transitions.
Sec.~\ref{Sec:Examples} is devoted to the numerical study of the two paradigmatic examples mentioned above.
Finally, in Sec.~\ref{Sec:Conclusion} we draw our conclusions and discuss possible perspectives for future studies.
In App.~\ref{App:Demonstrations} we include the proofs of some useful lemmas, while in App.~\ref{app:jordan} we consider an exactly-solvable model presenting a nondiagonalizable Liouvillian.


\section{Theoretical framework}
\label{subsec:proprieties}

In this work, we will consider open quantum systems where the coupling to a reservoir leads to a Markovian dynamics for the system density matrix $\hat{\rho} (t)$, described by a master equation in the Lindblad form \cite{BreuerBook07}
\begin{equation}\label{Eq:lindblad}
\partial_t \hat{\rho} (t)= -\frac{\ii }{\hbar} \left[\hat{H}, \hat{\rho} (t) \right ] + \sum_i \frac{\gamma_i}{2} \mathcal{D}[\hat{\Gamma}_i] \hat{\rho}(t),
\end{equation}
where $\hat{H}$ is the Hamiltonian describing the unitary evolution of the system, while the dissipation superoperators $\mathcal{D}[\hat{\Gamma}_i]$ are defined as
\begin{equation}\label{Eq:dissipator}
\mathcal{D}[\hat{\Gamma}_i] \hat{\rho}(t)= \left(2 \hat{\Gamma}_i \hat{\rho}(t) \hat{\Gamma}_i^\dagger - \hat{\Gamma}_i^\dagger  \hat{\Gamma}_i \hat{\rho}(t) - \hat{\rho}(t) \hat{\Gamma}_i^\dagger  \hat{\Gamma}_i\right).
\end{equation}
Each quantum jump operator $\hat{\Gamma}_i$ is associated with a dissipation channel occurring at the rate $\gamma_i$.
In the following, we will consider the case where $\hat{H}$,  $\hat{\Gamma}_i$, and $\gamma_i$ are time independent (for each $i$).
This kind of master equation can be applied for example to photonic quantum systems (see, for example, Refs. \cite{HouckNatPhys12,FitzpatrickPRX17,RodriguezPRL17,FinkPRX17,FinkNATPHY18}).

Since the Lindblad master equation~\eqref{Eq:lindblad} is  linear in $\hat{\rho}$, it is possible to associate with it the so-called Liouvillian superoperator $\L$, defined as
\begin{equation}\label{eq:liouvillian}
\partial_t \hat{\rho} (t)=\L \hat{\rho}(t).
\end{equation}
The superoperator $\L$ of the Lindblad master equation is trace-preserving and generates a completely positive map $e^{\L t}$ describing the time evolution of the system \cite{CarmichaelBOOK98,GardinerBOOK04,HarocheBOOK06,RivasBOOK11}.
For a time-independent Liouvillian, there is at least one steady state (if the dimension of the Hilbert space is finite \cite{RivasBOOK11}), i.e., a matrix such that 
\begin{equation}
\L \sss = 0. 
\end{equation}
This equation means that the steady-state density matrix is an eigenmatrix of the superoperator $\L$ corresponding to the zero eigenvalue. 
Moreover, under quite general conditions (see Refs.~\cite{AlbertPRA14,NigroarXiv18} and the App.~\ref{App:DiagonalisableLiouvillian}), the steady state is unique.
As we will see below, dissipative phase transitions are strictly related to the violation of this unicity condition.

Let us call $\mathcal{H}$ the Hilbert space of the system.
A density matrix $\hat\rho$, as any other operator $\hat{\xi}$, belongs to the operator space $\mathcal{H} \otimes \mathcal{H}$.
The Liouvillian superoperator, instead, is $\mathcal{L}\in L=\left(\mathcal{H} \otimes \mathcal{H}\right)^* \otimes \left(\mathcal{H} \otimes \mathcal{H}\right)$, where $L$ is the Liouville space.
In this article, we will systematically adopt the following notation: operators will be denoted by hats (e.g.,  $\hat{A}$),  superoperators will be written in calligraphic characters (e.g., $\mathcal{A}$), and states and their duals will be expressed in the Dirac notation ($\ket{a}$ and $\bra{a}$). 
A vectorized representation of an operator $\hat{A}$ will be denoted by $\vec{A}$, while the matrix representing a superoperator $\mathcal{A}$ is indicated by $\bar{\bar{\mathcal{A}}}$.
In particular, the matrix representation of the Liouvillian is:
\begin{equation}
	\begin{split}
	\supmat{\mathcal{L}} 
	=& -\ii \left((\hat{H}\otimes \mathds{1})  - (\mathds{1}\otimes \hat{H}^{\rm TR})\right) \\ & \quad+ \frac{\gamma}{2} \left(2 \hat{\Gamma}\otimes \hat{\Gamma}^*-\hat{\Gamma}^\dagger \hat{\Gamma}\otimes \mathds{1} -  \mathds{1} \otimes  \hat{\Gamma}^{\rm TR} \hat{\Gamma}^* \right),
	\end{split}
\end{equation} 
where the superscript {\rm TR} denotes the transposition.
Moreover, we will introduce the Hilbert-Schmidt inner product
\begin{equation}
\langle\hat{A},\hat{B}\rangle = \Tr{\hat{A}^\dagger \hat{B}},
\end{equation}
which, in the vectorized representation, takes the intuitive form $\vec{A}\cdot\vec{B}=\sum_{\mu, \nu} (\hat{A}^*)_{\mu,\nu}(\hat{B})_{\mu,\nu}$.
The definition of the norm naturally follows as
\begin{equation}
\|\hat{A}\|^2=\Tr{\hat{A}^\dagger \hat{A}}=\vec{A} \cdot \vec{A}=\sum_{\mu, \nu} |\hat{A}_{\mu,\nu}|^2.
\end{equation}

\subsection{Spectral properties  of Liouvillian superoperators}\label{subsec:eigenvalues}

To fully determine the dynamics of the system, the knowledge of the steady-state density matrix $\sss$ is not enough. Indeed, one has to know all the spectrum of the Liouvillian superoperator $\L$, whose
eigenmatrices  and eigenvalues are defined via the relation 
\begin{equation}\label{eq:Liouvillian_Eigenstates}
\mathcal{L} \eig{i} = \lambda_i \eig{i}.
\end{equation}
Equivalently, in the vector-representation, $\vec{\rho}_{i}$ is a right-eigenvector of the superoperator matrix $\supmat{\L}$.
Having introduced a norm, we require the eigenstates to be normalized: $\|\eig{i}\|^2=1$ \footnote{The steady-state density matrix $\sss$ is thus proportional to the eigenstate of $\L$ whose eigenvalue is zero, since $\sss$ must satisfy $\Tr{\sss}=1$, which may not correspond to a state with norm one.}.
Since $\L$ is not Hermitian, its eigenvectors are, in general, not orthogonal: $\vec{\rho}_i\cdot\vec{\rho}_j\neq0$.
If the Liouvillian is diagonalizable, we can conveniently use the eigenstates of $\L$ as a basis of the Liouville space, apart from some exceptional points (see App.~\ref{App:DiagonalisableLiouvillian} and Ref. \cite{MacieszczakPRL16}).
Under this hypothesis, for any operator $\hat{A}$ there exists a unique decomposition
\begin{equation}
\label{Eq:eigenvectordecomposition}
\hat{A}=\sum_i c_i \eig{i}.
\end{equation}

It can be proved \cite{BreuerBook07,RivasBOOK11} that $\Re{\lambda_i}\leq 0, \forall i$.
Since the real part of the eigenvalues is responsible for the relaxation to the steady-state,
$\sss = \lim_{t \to + \infty} e^{{\mathcal L} t} \rho(0)$.
For convenience, we sort the eigenvalues in such a way that $\abs{\Re{\lambda_0}}<\abs{\Re{\lambda_1}} < \ldots < \abs{\Re{\lambda_n}}$. 
From this definition it follows that $\lambda_0=0$ and $\sss=\eig{0}/\Tr{\eig{0}}$.
We can also identify another relevant quantity: the Liouvillian gap $\lambda=\abs{\Re{\lambda_1}}$, which is also called the asymptotic decay rate \cite{KesslerPRA12}, determining the slowest relaxation dynamics in the long-time limit. 

For any Liouvillian, the following lemmas hold (for a detailed proof of Lemmas 3 and 4 see App.~\ref{App:ProvesOnLiouvillian}):
\begin{itemize}
	\item{\bf Lemma 1.} Given Eq.\eqref{eq:Liouvillian_Eigenstates}, $e^{\L t}\eig{i}=e^{\lambda_i t}\eig{i}$ .
	\item{\bf Lemma 2.} $\Tr{\eig{i}}=0$ if $\Re{\lambda_i} \neq 0$. \\
	Indeed, the Liouvillian evolution conserves the trace \cite{CarmichaelBOOK98,GardinerBOOK04,HarocheBOOK06,RivasBOOK11} and if $\Re{\lambda_i} \neq 0$  for $t \to + \infty$ we have $e^{\L t}\eig{i}=e^{\lambda_i t}\eig{i} \to 0$. 
	\item{\bf Lemma 3.} If $\L\eig{i}=\lambda_i\eig{i}$ then  $\L\eig{i}^\dagger=\lambda_i^*\eig{i}^\dagger$.\\
	This implies that, if $\eig{i}$ is Hermitian, then $\lambda_i$ has to be real.
	Conversely, if $\lambda_i$ is real and of degeneracy 1, $\eig{i}$ is Hermitian.
	If $\lambda_i$ has geometric multiplicity $n$ and $\L$ is diagonalizable, it is always possible to construct $n$ Hermitian eigenmatrices of $\L$ with eigenvalue $\lambda_i$ \footnote{The algebraic multiplicity of $\lambda$ is defined as the number of times $\lambda$ appears as a root of the characteristic equation.
		The geometric multiplicity, instead, is the maximum number of linearly independent eigenvectors associated with $\lambda$.}.
	\item{\bf Lemma 4.}  If $\lambda_i=0$ has degeneracy $n$, then there exist $n$ independent eigenvectors of the Liouvillian (the algebraic multiplicity is identical to the geometrical one). Therefore, there exist $n$ different steady states towards which the system can evolve, depending on the initial condition.
\end{itemize}

\subsection{Spectral decomposition of density matrices}

Let us consider a system admitting a unique steady state.
To be physical, its $\hat{\rho}(t)$ must be a Hermitian, positive-definite matrix with trace equal to one.
Hence, from Lemma~2, to ensure $\Tr{\rho(t)}=1$ at every time, we must have:
\begin{equation}
\label{eq:time_evol}
\hat{\rho}(t)=\frac{\eig{0}}{\Tr{\eig{0}}}+ \sum_{i\neq 0} c_i (t) \eig{i}=\sss + \sum_{i\neq 0} c_i(0) e^{\lambda_i t} \eig{i}.
\end{equation}

\subsubsection{The case of a real Liouvillian eigenvalue $\lambda_i$}
When $\lambda_i$ is real, $\eig{i}$ can be constructed to be Hermitian (see Lemma~3 of Sec.~\ref{subsec:eigenvalues}).
Thus, it can be diagonalized, obtaining the spectral decomposition \cite{RivasBOOK11}
\begin{equation}
\eig{i}= \sum_n p_n^{(i)} \ket{\psi_n^{(i)}}\bra{\psi_n^{(i)}},
\end{equation}
where all the $p_n^{(i)}$ must be real and $\braket{\psi_n^{(i)}|\psi_m^{(i)}}=\delta_{n,m}$.
Moreover, since $\eig{i}$ is traceless (see Lemma~2 of Sec.~\ref{subsec:eigenvalues}), some of the $p_n^{(i)}$ must be positive and the others negative.
We can order them in such a way to have $p_n^{(i)}>0$ for $n\leq\bar{n}$ and $p_n^{(i)}<0$ for $n> \bar{n}$.
Thus, we have:
\begin{equation}\label{Eq:DecoRhoReal}
\hat{\rho}_i \propto \hat{\rho}_i^+ - \hat{\rho}_i^-,
\end{equation}
where
\begin{eqnarray}
\hat{\rho}_i^+&=&\sum_{n\leq\bar{n}} p_n^{(i)} \ket{\psi_n^{(i)}}\bra{\psi_n^{(i)}} , \cr
&&\cr
\hat{\rho}_i^- &=& - \sum_{n>\bar{n}} p_n^{(i)} \ket{\psi_n^{(i)}}\bra{\psi_n^{(i)}}
\end{eqnarray}
and where the $\{p_n\}$ have been normalized to ensure $\Tr{\hat{\rho}_i^+}=\Tr{\hat{\rho}_i^-}=1$.
With this definition, $\eig{i}^\pm$ are density matrices.
Consequently, a state of the form $\hat{\rho}(0)=\sss+A \, \eig{i}$ will evolve in time as \cite{MacieszczakPRL16}
\begin{equation}
\hat{\rho}(t)=\sss+A e^{\lambda_i t}(\hat{\rho}_i^+ - \hat{\rho}_i^-).
\end{equation}
\\
\subsubsection{The case of a complex Liouvillian eigenvalue $\lambda_i$}
Let us now consider an eigenmatrix $\eig{i}$ with a complex eigenvalue $\lambda_i$.
As it stems from Eq.~\eqref{eq:time_evol}, to ensure an Hermitian $\hat{\rho}(t)$ such an eigenmatrix must always appear in combination with its Hermitian conjugate $\eig{i}^\dagger$, which is also an eigenmatrix of $\L$ (Lemma~3 of Sec.~\ref{subsec:eigenvalues}).
Thus, one can simply consider the Hermitian combinations $\eig{i}+\eig{i}^\dagger$ and $\ii \, (\eig{i}-\eig{i}^\dagger)$.
For example, given an initial condition $\hat{\rho}(0)=\sss+A(\eig{i}+\eig{i}^\dagger)$ with $A$ real, one has \cite{MacieszczakPRL16}:
\begin{widetext}
\begin{equation}
\begin{split}
\hat{\rho}(t)&=\sss+A\left(e^{\lambda_i t}\eig{i}+e^{\lambda_i^* t}\eig{i}^\dagger\right)=
\sss +A\, e^{\Re{\lambda_i} t} \left( \eig{i} e^{\ii \Im{\lambda_i} t} + \eig{i}^\dagger e^{-\ii \Im{\lambda_i} t} \right)\\
&= \sss +2 A\, e^{\Re{\lambda_i} t} \left[\left( \eig{i} + \eig{i}^\dagger \right) \cos(\Im{\lambda_i} t) + \ii \left( \eig{i} - \eig{i}^\dagger \right) \sin(\Im{\lambda_i} t) \right].
\end{split}
\end{equation}
\end{widetext}

\subsection{Definition of dissipative phase transitions}
\label{Sec:Phase_transitions}

Let us consider a system where a thermodynamic limit is obtained when a parameter $N \to + \infty$.
For example, in a lattice of spins, $N$ would be the number of lattice sites.
For any finite $N$, the system always admits a unique steady-state solution.
In the thermodynamic limit $N\to+\infty$, a transition between two different phases is characterized by the nonanalytical behavior of some $\zeta$-independent observable $\hat{o}$ when the parameter $\zeta$ tends to the critical value $\zeta_c$.
Formally, we say that there is a phase transition of order $M$ if
\begin{equation}\label{Eq:DPTDefinition}
 \lim_{\zeta \to  \zeta_c}\left|\frac{\partial^M}{\partial \zeta^M} \lim_{N\rightarrow +\infty}\Tr{\sss(\zeta, N) \hat{o}}\right|=+ \infty.
\end{equation}
Since $\hat{o}$ does not depend on $\zeta$, the discontinuity in Eq.~\eqref{Eq:DPTDefinition} is due to a discontinuous behavior in $\sss(\zeta,N\to\infty)$.
As proved in \cite{KatoBOOK}, a discontinuity of an eigenmatrix is to be associated with a level crossing in the spectrum of the Liouvillian. Since $\sss$ is associated with $\lambda_0=0$, the phase transition must coincide with the closure of the Liouvillian gap \cite{KesslerPRA12,HorstmannPRA2013} (indeed, in this case, is more correct to talk about {\it level touching}).
Therefore, dissipative phase transitions are intimately connected to the emergence of multiple steady states in the thermodynamic limit $N\to+\infty$.


\section{First order Phase transition}
\label{Sec:First_order_Phase_transition}

\begin{figure}[t!]
	\centering
	\includegraphics[width=0.8\linewidth]{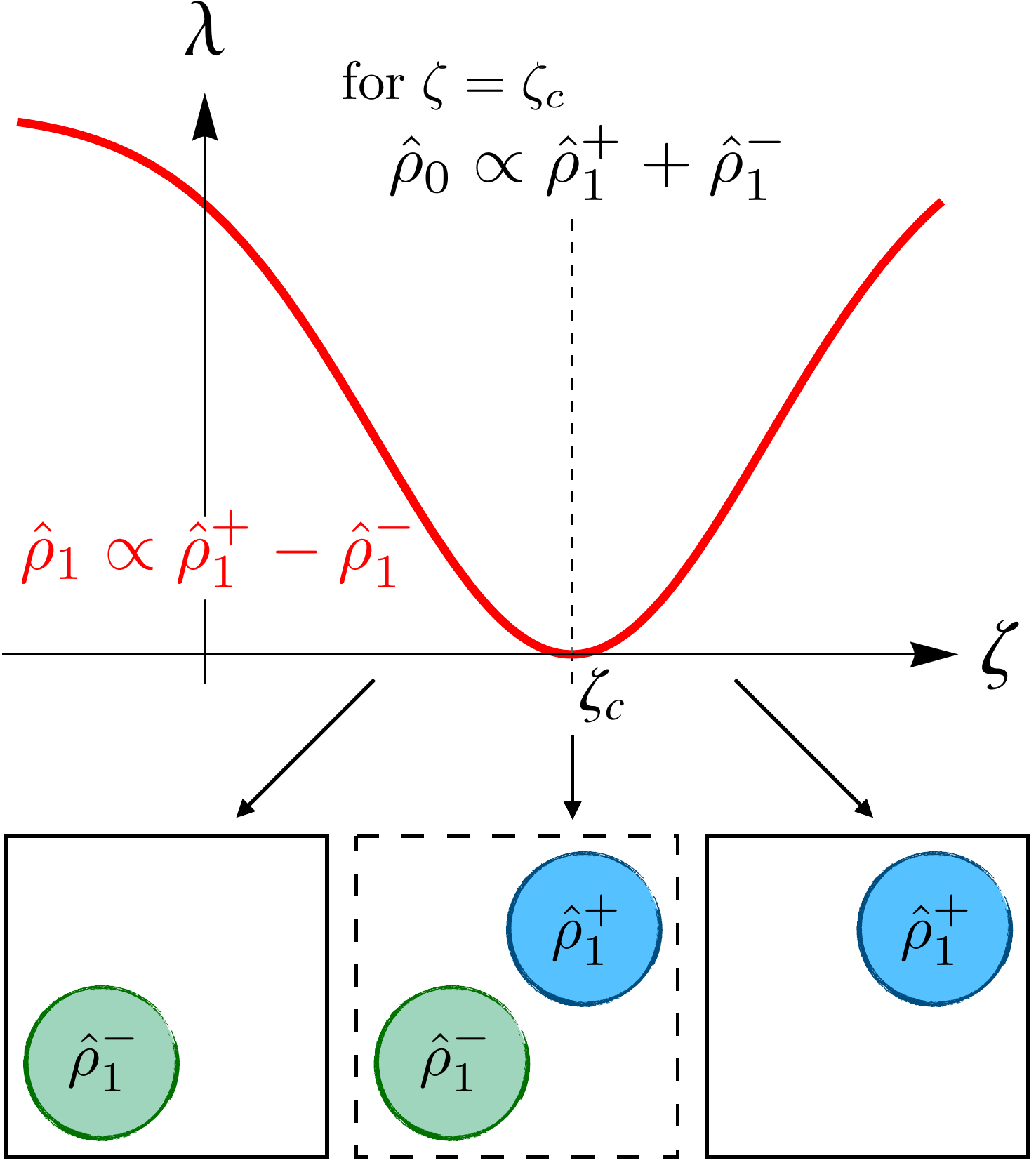}	
	\caption{
	Sketch depicting the paradigm of a first-order dissipative phase transition, formally described in Sec.~\ref{Sec:First_order_Phase_transition}.
	In the thermodynamic limit, the Liouvillian gap $\lambda=\left|\Re{\lambda_{1}}\right|$ closes when the parameter $\zeta$ of the Liouvillian assumes the critical value $\zeta_c$.
	We note that, for $\zeta\simeq\zeta_c$, we must also have $\Im{\lambda_{1}}=0$.
	Right before (after) the critical point, the steady-state density matrix $\sss\simeq\hat{\rho}_1^-$ ($\sss\simeq\hat{\rho}_1^+$),
	which represents one of the two different phases of the system.
	At the critical point $\zeta=\zeta_c$, $\sss$ is bimodal: the steady state is a statistical mixture of $\hat{\rho}_1^+$ and $\hat{\rho}_1^-$.}
	\label{fig:Scheme1}
\end{figure}

In this section we consider the emergence of a first-order dissipative phase transition at $\zeta=\zeta_c$ in the thermodynamic limit $N\to+\infty$. 
Such a transition must be associated with the existence of two different steady states, one for $\zeta<\zeta_c$ and the other for $\zeta>\zeta_c$, which implies that
\begin{equation}
\lambda_{1}(\zeta,N\to+\infty)\neq\lambda_0=0\quad{\rm for}\quad\zeta\neq\zeta_c.
\end{equation}
According to our definition, a first-order dissipative phase transition occurs when Eq.~\eqref{Eq:DPTDefinition} is satisfied for $M=1$, which also corresponds to
\begin{equation}
\label{eq:rhossfo}
\lim_{\zeta \to  \zeta_c^+}  \lim_{N\to + \infty} \sss(\zeta,N)\equiv\hat{\rho}^+ \neq
\hat{\rho}^- \equiv  \lim_{\zeta \to  \zeta_c^-}  \lim_{N\to + \infty} \sss(\zeta,N),
\end{equation}
which defines $\rho^+$ ($\rho^-$) as the steady state in the thermodynamic limit right after (before) the critical point.
From Eq. \eqref{eq:rhossfo} we can write  that $\sss(\zeta) = \theta(\zeta-\zeta_c)\hat\rho^+ + \theta(\zeta_c-\zeta)\hat\rho^-$ for $\zeta\neq\zeta_c$, where $\theta(x)$ is the Heaviside step function.
Assuming the continuity of the Liouvillian, we can state that $\L(\zeta_c)\hat{\rho}^\pm=0$ (we drop the explicit dependence on $N$ when assuming the thermodynamic limit).
This implies that $\lambda_1(\zeta_c)=\lambda_0=0$ and hence $\sss(\zeta_c)$ and $\eig{1}(\zeta_c)$ belong to the kernel spanned by $ \hat{\rho}^\pm$.
It is worth stressing that in the thermodynamic limit and for $\zeta=\zeta_c$, \emph{both} the real and imaginary part of $\lambda_{1}$ must vanish.
Furthermore, in a first-order dissipative phase transition, the condition $\Im{\lambda_{1}}=0$ must hold in a finite domain around $\zeta_c$, as a consequence of Lemma 3
\footnote{
		Suppose that for $\zeta\simeq\zeta_c$, $\Im{\lambda_{1}(\zeta)}\neq0$.
		In virtue of Lemma 3, also $\lambda_2(\zeta)=\lambda_1^*(\zeta)$ belongs to the spectrum of $\L$.
		If, in the thermodynamic limit, $\Im{\lambda_{1}(\zeta)}$ goes to zero only at $\zeta=\zeta_c$, so does $\Im{\lambda_{2}(\zeta)}$.
		This would result in having three zeros in the Liouvillian spectrum at $\zeta=\zeta_c$.
		This is in contrast with the present theory of first-order dissipative phase transitions, which predicts a double degeneracy.
		Hence, the imaginary part of $\lambda_{1}$ must be zero in a finite domain around $\zeta_c$
}.

Lemma~2 (Sec.~\ref{subsec:proprieties}) ensures that $\Tr{\eig{1}(\zeta)}=0$ if $\lambda_1(\zeta)\neq0$ (i.e., $\zeta\neq\zeta_c$). Moreover, as discussed in \cite{KatoBOOK}, $\lambda_1(\zeta)$ must be continuous in a domain of the parameter space around $\zeta=\zeta_c$ (c.f. Fig.~\ref{fig:Scheme1}).
By analogy, we want also $\eig{1}(\zeta)$ to be continuous around $\zeta_c$,
and to extend the zero-trace condition we must set
\begin{equation}
\label{eq:rho1fo}
\eig{1}(\zeta_c) \propto  \hat{\rho}^+ - \hat{\rho}^-.
\end{equation}
The above equation allows the identification of the states $\eig{1}^\pm$ obtained with the eigendecomposition~\eqref{Eq:DecoRhoReal} with the two phases $\hat{\rho}^\pm$ [Eq.~\eqref{eq:rhossfo}] emerging  in the thermodynamic limit.
Together with the continuity requirement, this allows ud to interpret $\eig{1}^\pm(\zeta)\simeq\hat{\rho}^\pm$ in a domain around $\zeta=\zeta_c$. 
In this region, since the Liouvillian gap is finite, we also have $\eig{0}(\zeta)\propto\sss(\zeta)$. Using that $\theta(0)=1/2$, we can infer 
\begin{equation}
\label{eq:rho0fo}
\eig{0}(\zeta_c) \propto \hat{\rho}^+ + \hat{\rho}^-.
\end{equation}
Accordingly,  $\eig{0}(\zeta_c)$ and $\eig{1}(\zeta_c)$ are orthogonal, since $\braket{\eig{0}(\zeta_c), \eig{1}(\zeta_c)} \propto \Tr{\left(\rho^+\right)^2} -  \Tr{\left(\rho^-\right)^2} = 0$.

For large but finite $N$, provided that $|\Re{\lambda_2}|\gg|\Re{\lambda_1}|>0$, Eqs.~\eqref{eq:rho1fo} and~\eqref{eq:rho0fo} are asymptotic good approximations and, since $\hat\rho^\pm=\eig{1}^\pm(\zeta_c,N)$, we get the asymptotic expression
\begin{equation}
\label{eq:rhosfo}
\sss(\zeta_c,N) \simeq \frac{\eig{1}^+(\zeta_c,N) + \eig{1}^-(\zeta_c,N)}{2},
\end{equation}
which ensures Hermiticity and unit trace of the $\sss(\zeta_c,N)$.
Let us note that Eq. \eqref{eq:rhosfo} has a clear physical interpretation: at the critical point, for a finite-size system, the steady state is the equiprobable mixture of the two phases, which are encoded in the spectral decomposition of $\eig{1}(\zeta_c,N)$. 
Remarkably, in a small region on the left (right) of the critical point, $\eig{1}^+$ ($\eig{1}^-$) is metastable.
This means that if the system is initialized in one of these two states it will remain stuck, for a time proportional to $1/\lambda$, before reaching the steady-state \cite{MacieszczakPRL16}.
This can give rise to hysterical behavior, typical of first-order phase transitions \cite{RodriguezPRL17}.

Conversely, if $\lambda_{1}=0$ in a point, one has to have a first-order phase transition.
A proof can be found in App.~\ref{app:first_order_gap_demonstration}.


\section{Second order phase transitions with symmetry breaking}
\label{Sec:Secon_order}

\begin{figure}[t!]
	\centering
	\includegraphics[width=0.8\linewidth]{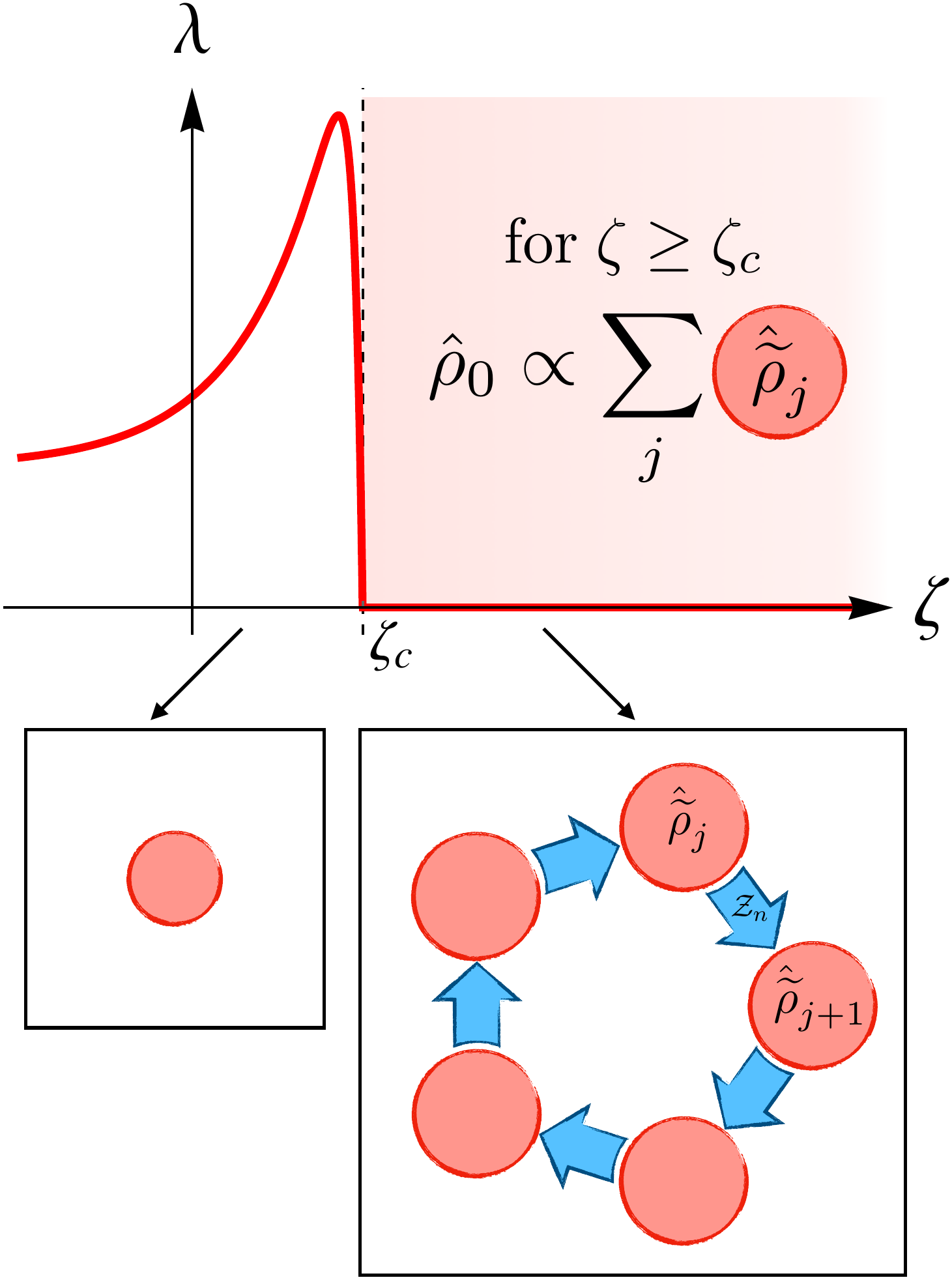}	
	\caption{
	Sketch depicting the paradigm of a second-order dissipative phase transition (cf. Sec.~\ref{Sec:Secon_order}), associated with the breaking of a $\mathcal{Z}_n$ symmetry (in the sketch $n=5$).
	In the thermodynamic limit, the Liouvillian gap $\lambda$ closes over the whole region $\zeta\geq \zeta_c$, $\zeta$ being the critical parameter triggering the transition.
	Moreover, one has that $\lambda_0,\cdots,\lambda_{n-1}=0$ for $\zeta\geq \zeta_c$.
	When $\lambda\neq0$ (here for $\zeta<\zeta_c$), the steady-state density matrix $\hat{\rho}_{ss}$ is mono-modal.
	In the symmetry-broken phase ($\lambda=0$ and $\zeta\geq \zeta_c$), $\hat{\rho}_{ss}$ is an $n$-modal statistical mixture of density matrices $\hat{\widetilde{\rho}}_j$, which are mapped one onto the other under the action of the symmetry superoperator $\mathcal{Z}_n$.}
	\label{fig:Scheme2}
\end{figure}

In this section, we will consider second-order dissipative phase transitions associated with a symmetry breaking.
A symmetry of an open quantum system is described by a unitary superoperator $\mathcal{U}=\hat{V}\bigcdot \hat{V}^{-1}$
(where $\hat{V}$ is a unitary operator and the $\bigcdot$ in the previous definition means that, upon the action of the superoperator $\mathcal{U}$ on a generic operator, the latter has to be inserted in between $\hat{V}$ and $\hat{V}^{-1}$)
\cite{BaumgartnerJPA08}, such that
\begin{equation}
\mathcal{U}^{-1}\mathcal{L}\,\mathcal{U}=\mathcal{L},
\end{equation}
or, equivalently, $[\mathcal{L},\mathcal{U}]=0$. 
It follows that the matrix representations $\supmat{\mathcal{U}}$ of $\mathcal{U}$ and $\supmat{\L}$ of $\mathcal{L}$ can be simultaneously diagonalized.
From now on, we will call the symmetry sector $L_u$ the subspace of the Liouville space $L$ spanned by the eigenmatrices of $\mathcal{U}$ with eigenvalue $u$. 
The existence of a symmetry means that the Lindblad master equation cannot mix different symmetry sectors. Therefore $\supmat{\L}$  can be cast in a block-diagonal form:
\begin{equation}
\label{Eq:block_diagonal}
\supmat{\L}=
\left[
\begin{array}{c c c c}
\supmat{\L}_{u_0} & 0 & \dots & 0 \\
0& \supmat{\L}_{u_1} & \dots & 0 \\
\vdots & \vdots& \ddots & \vdots \\
0 & 0 & \dots & \supmat{\L}_{u_n}
\end{array}
\right].
\end{equation}

Consider an arbitrary density matrix $\hat{\rho}$ which is an eigenmatrix of $\mathcal{U}$: $\mathcal{U} \hat{\rho}= u \hat{\rho}$.
Taking the trace of both  sides of the previous identity, and given the form of $\mathcal{U}$, one finds $u=1$.
If $\sss$ is the only eigenmatrix with zero eigenvalue of $\L$ (unique steady state), it must also be an eigenmatrix of $\mathcal{U}$.
From a physical perspective, this tells us that the symmetry sector to which $\sss$ (and therefore $\eig{0}$) belongs is \emph{always} $L_{u=1}$.

A symmetry-breaking dissipative phase transition is associated with the emergence of multiple eigenmatrices of $\L$ with $\lambda_i=0$, each of them belonging to a different symmetry sector $L_{u_i}$.
The structure imposed by Eq.~\eqref{Eq:block_diagonal} is preserved and the previous considerations still hold.
Therefore, $\eig{0}$ (belonging to the symmetry sector $L_{u=1}$) is still the only eigenmatrix of $\L$ with nonzero trace.

The block-diagonal structure of the Liouvillian (see Eq.~\eqref{Eq:block_diagonal}), together with the previous observations, can play a fundamental role in reducing the complexity of the problem. Indeed, by properly exploiting spatial and/or internal symmetries, one can explicitly construct the reduced subspace in which the steady-state density matrix belongs. This can give a substantial speed-up for algorithms based on Monte Carlo strategies \cite{SavonaARX18}, cluster expansions \cite{BiellaPRB2018,BiondiNJP2018}, corner methods \cite{FinazziPRL15}, and the tensor-network ansatz \cite{OrusNatComm17,VidalPRL07,ZwolakPRL04}.

\subsubsection{$\mathcal{Z}_2$ symmetry }

Let us consider first a system which has a discrete $Z_2$ symmetry represented by the superoperator ${\mathcal{Z}_2 = \hat{Z}_2 \bigcdot \hat{Z}_2^\dagger}$.
Later, we will deal with the general case of a $Z_n$ symmetry.
The symmetry superoperator $\mathcal{Z}_2$ admits two eigenvalues, namely $\pm 1$. 
For $\zeta<\zeta_c$ ($\zeta$ being the critical parameter) there exists a unique steady state associated with the eigenvalue $\lambda_0=0$, and
$\mathcal{Z}_2 \sss = \sss$. 
For $\zeta \geq \zeta_c$, a phase transition with a symmetry breaking takes place.
Consequently, $\lambda_0=\lambda_1=0$ while $\eig{0}$ and $\eig{1}$ belong to two different symmetry sectors (cf. Fig.~\ref{fig:Scheme2}).
From these properties, it follows that $\eig{0}$ and $\eig{1}$ are orthogonal, since
\begin{equation}
\label{eq:orthogonal_eigenlatricers_Z2}
\braket{\eig{0}| \eig{1}} =\braket{\mathcal{Z}_2\eig{0}| \eig{1}} =\braket{\eig{0}| \mathcal{Z}_2\eig{1}} = -\braket{\eig{0}| \eig{1}},
\end{equation}
where we exploited the Hermiticity of $\mathcal{Z}_2$.
Similarly, $\braket{\eig{0}^\dagger| \eig{1}}=0$.
Since $\lambda_1= 0$ is real and $\eig{1}^\dagger \neq \eig{0}$, the eigenmatrix $\eig{1}$ is Hermitian (lemmas of Sec.~\ref{subsec:eigenvalues}). 
Hence, the density matrices 
\begin{equation}
\hat{\rho}^\pm =  \frac{\eig{0} \pm \eig{1}}{\Tr{\eig{0}}} 
\end{equation}
 are steady states of the master equation breaking the symmetry, as
$\mathcal{Z}_2 \hat{\rho}^\pm =\hat{\rho}^\mp$.
From Eq.~\eqref{eq:orthogonal_eigenlatricers_Z2} it follows that $\hat{\rho}^+$ and $\hat{\rho}^-$ are orthogonal as well.
So that we have:
\begin{subequations}
	\begin{align}
	 \eig{0} &\propto \hat{\rho}^+ + \hat{\rho}^-,\\
	 \eig{1} &\propto \hat{\rho}^+ - \hat{\rho}^-.
	\end{align}		
\end{subequations}
Thus, we can conclude that the two symmetry-broken states $\hat{\rho}^\pm$  are the two matrices stemming from the spectral decomposition of $\eig{1}$, i.e., $\eig{1}^\pm$ [c.f. Eq~\eqref{Eq:DecoRhoReal}].
For a finite-size system, where the steady state is unique, 
\begin{equation}
\sss(\zeta\geq\zeta_c,N) \simeq \frac{\eig{1}^+(\zeta_c,N) + \eig{1}^-(\zeta_c,N)}{2}.
\end{equation}

Since we are considering a second-order phase transition, we must ensure that the unique steady state in $\zeta_c^-$ coincides with both the symmetry-breaking steady states in $\zeta_c^+$: $\sss(\zeta_c^-)=\hat{\rho}^+(\zeta_c^+)=\hat{\rho}^-(\zeta_c^+)$.
Consequently, according to this discussion, $\eig{1}(\zeta_c)=0$.
Therefore, a second-order phase transition is characterized by the coalescence of two eigenvectors of the Liouvillian, which may give rise to a Jordan form of the Liouvillian (see App.~\ref{app:jordan}).
In order to unveil the symmetry breaking in a finite-size system (where the symmetry is always preserved) one can resort to different strategies. To identify the critical point, one can use an external weak probe which breaks the symmetry (see for example Refs. \cite{RotaPRB17,BiellaPRB2018}) and look for divergences in the associated susceptibility.
To characterize the existence of the two (or more) metastable states which individually break the symmetry, one can also resort to a quantum trajectory protocol \cite{Daley}.
Indeed, the dynamics of a single trajectory can explicitly break the symmetry, even if once the average over many trajectories is taken, such a symmetry is restored \cite{BartoloEPJST17,RotaNJP18}.

\subsubsection{$\mathcal{Z}_n$ symmetry }

Consider now a generic symmetry superoperator ${\mathcal{Z}_n = \hat{Z}_n \bigcdot \hat{Z}_n^\dagger}$.
In this case, the Liouvillian can be partitioned into $n$ blocks, each characterised by an eigenvalue $z_j=\exp[2 \, \ii \, \pi j /n]$, with $j=0,1 \ldots n-1$  (i.e. the eigenvalues must satisfy the equation $z_j^n=1$).
In the symmetry-broken phase, in each of those blocks there exists an eigenmatrix $\eig{j}$ such that $\L \eig{j} =0$ and $\mathcal{Z}_n \eig{j} = z_j \eig{j}$.
Lemma~3 of Sec.~\ref{subsec:eigenvalues} imposes $\L\eig{j}^\dagger=0$.
Moreover,  $\eig{j}^\dagger$ is also an eigenmatrix of $\mathcal{Z}_n$ of eigenvalue $z_j^*$, since
\begin{equation}
\mathcal{Z}_n \eig{j}^\dagger =\left(\hat{Z}_n \eig{j}^\dagger \hat{Z}_n^\dagger \right)=\left(\hat{Z}_n \eig{j} \hat{Z}_n^\dagger \right)^\dagger=z_j^* \eig{j}^\dagger.
\end{equation}
Note that, by definition $z_j^*=z_{n-j}$, and hence $\eig{j}^\dagger = \eig{n-j}$.
As a particular case, if $z_j=z_j^*$ then $\eig{j}=\eig{j}^\dagger$. 

To construct a basis of the degenerate subspace made of density matrices, consider the operator
\begin{equation}
\begin{split}
\symmat{0} &= \sum_{j=0}^{n-1} \frac{\eig{j}}{\Tr{\eig{0}}}=\sum_{j=0}^{n-1} \frac{\eig{j}+\eig{n-j}}{2 \, \Tr{\eig{0}}} =\sum_{j=0}^{n-1} \frac{\eig{j}+\eig{j}^\dagger}{2 \, \Tr{\eig{0}}}.
\end{split}
\end{equation}	
With this choice, $\symmat{0}$ is a density matrix, since it is Hermitian and it has trace 1 ($\Tr{\eig{j}}= \Tr{\eig{0}}\delta_{j,0}$).
For $\symmat{1}= \mathcal{Z}_n \symmat{0}$, one has
\begin{equation}
\symmat{1}=\sum_{j=0}^{n-1} \frac{z_j\eig{j}+z_j^*\eig{j}^\dagger}{2\, \Tr{\eig{0}}},
\end{equation}
which is still Hermitian and of unitary trace, and therefore a density matrix.
By iterative application of the symmetry operator $\mathcal{Z}_n$, and since $\symmat{i} \neq \symmat{j}$ for $i \neq j$, one obtains a basis $\{\symmat{i}\}$ of density matrices, with $i=0,\cdots, n -1$. 
In compact notation, one has
\begin{equation}
\label{eq:symbasis_construcion}
\symmat{l} = \mathcal{Z}_n^l \sum_{j=0}^{n-1} \frac{\eig{j}}{\Tr{\eig{0}}} = \sum_{j=0}^{n-1} \frac{z_j^l(i)\eig{j}}{\Tr{\eig{0}}}.
\end{equation}
Equation~\eqref{eq:symbasis_construcion} can be inverted to obtain $\eig{k}$ as a function of $\symmat{l}$:
\begin{equation}
\label{eq:thetrick}
\begin{split}
\sum_{l=0} ^{n-1} \left(z_k ^*\right) ^l \symmat{l}  & =\sum_{l=0}^{n-1} \sum_{j=0}^{n-1}  \frac{\left(z_k ^* z_j \right)^l\eig{j} }{ \Tr{\eig{0}}}=\sum_{l=0}^{n-1} \sum_{j=0}^{n-1} \frac{z_{j-k} ^l \eig{j}}{\Tr{\eig{0}}} \\
&=\frac{n}{\Tr{\eig{0}}} \sum_{j=0}^{n-1}  \delta_{j,k} \eig{j} = \frac{n}{\Tr{\eig{0}}}  \eig{k},
\end{split}
\end{equation}
where we used the identity
\begin{equation}
\sum_{l=0} ^{n-1} z_{j-k} ^l=\sum_{l=0}\left( e^{\frac{2 \ii \pi (j-k)}{n}}\right) ^l = n \, \delta_{k,j}.
\end{equation}
We conclude that 
\begin{equation}
\eig{k} \propto \sum_{l=0}^{n-1}\frac{(z_k^*)^l\symmat{l}}{n}.
\end{equation}
Summarizing, we have constructed a basis of $\{\symmat{i}\}$ of the kernel of the Liouvillian made of density matrices such  that $\mathcal{Z}_n \symmat{i} = \symmat{{\rm mod}(i+1,n)}$, as depicted in Fig.~\ref{fig:Scheme2}.
This construction ensures that $\mathcal{Z}_n\eig{0} = \eig{0}$.
Again, for large enough but finite $N$, where the steady state is unique also for $\zeta\geq\zeta_c$, we get the asymptotic expression 
\begin{equation}
\sss(\zeta\geq\zeta_c,N) \simeq \sum_{l=0}^{n-1}\frac{\symmat{l}(\zeta\geq\zeta_c,N)}{n}.
\end{equation}



\section{Applications to specific models}\label{Sec:Examples}

In the following, we will explore some specific models exhibiting dissipative phase transitions in a thermodynamic limit. We will show that, in the finite-size case, our theory predicts with high fidelity the form of $\sss$ in the vicinity of the critical point. In particular, we will analyze some systems for which a brute force diagonalization of the Liouvillian supermatrix $\supmat{\L}$ is possible.

\subsection{The driven-dissipative Kerr resonator}
\begin{figure}[t!]
	\centering
	\includegraphics[width=0.98\linewidth]{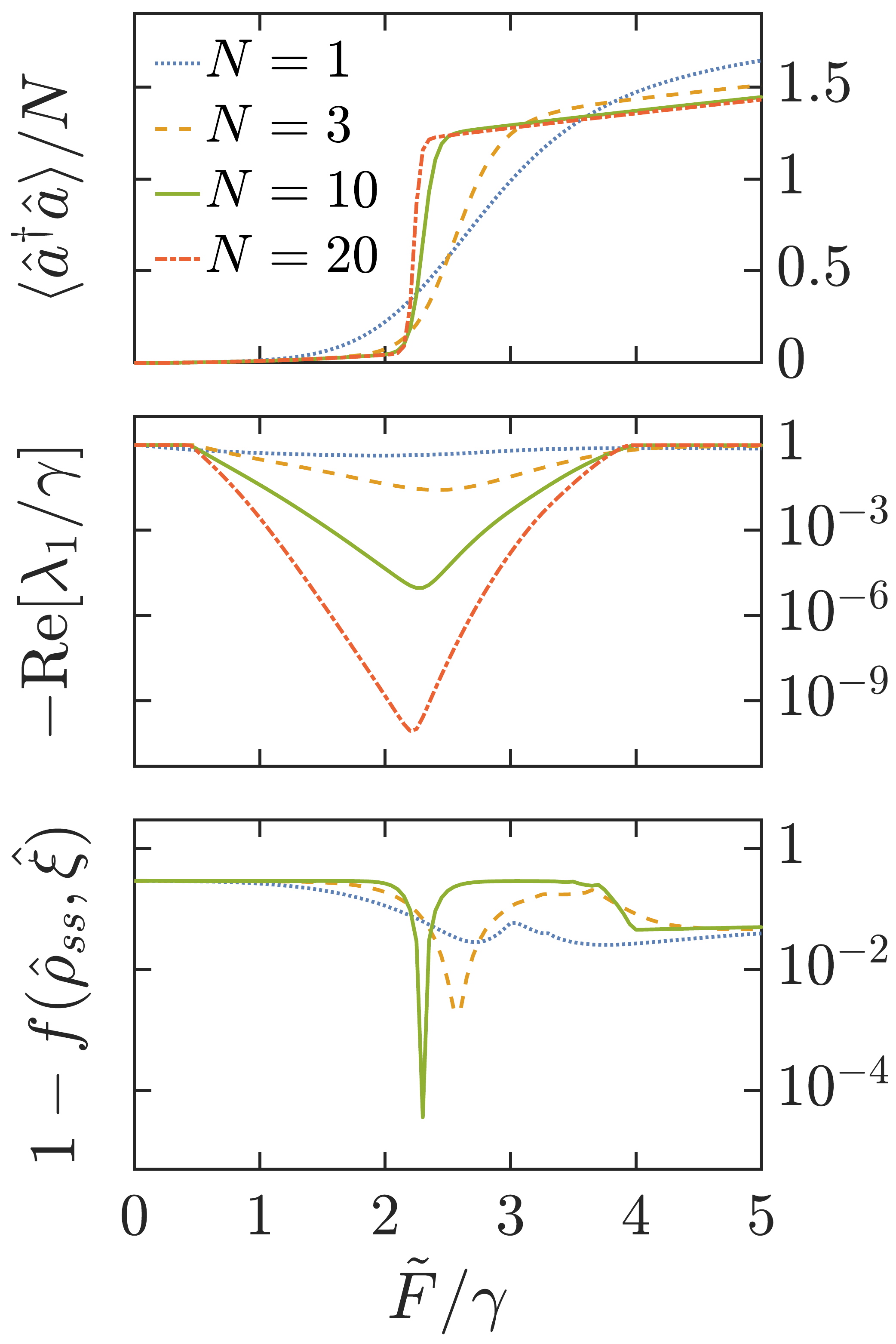}
	\caption{Numerical results for the driven-dissipative Kerr  model.
		Top panel: Rescaled number of photons $\braket{\hat{a}^\dagger \hat{a}}/N $ as a function of the rescaled driving $\tilde{F}/\gamma$ for different values of $N$.
		Middle panel: -${\rm Re} [\lambda_1/\gamma]$ (Liouvillian gap) for different values of $N$.
		In the selected range of parameters we find that $\Im{\lambda_1}$ is zero within the numerical error.
		Bottom panel: The error $1-f$, where $f$ is the fidelity between the steady-state density matrix $\sss$ and the one reconstructed via the eigendecomposition of the first eigenstate $\hat{\xi} = (\eig{1}^+ + \eig{1} ^-)/2$.
	    Parameters: $\Delta/\gamma=10$, $\tilde{U}/\gamma=10$.}
	\label{fig:Kerr_resonator_transition}
\end{figure}

\begin{figure}[t!]	
	\includegraphics[width=0.98\linewidth]{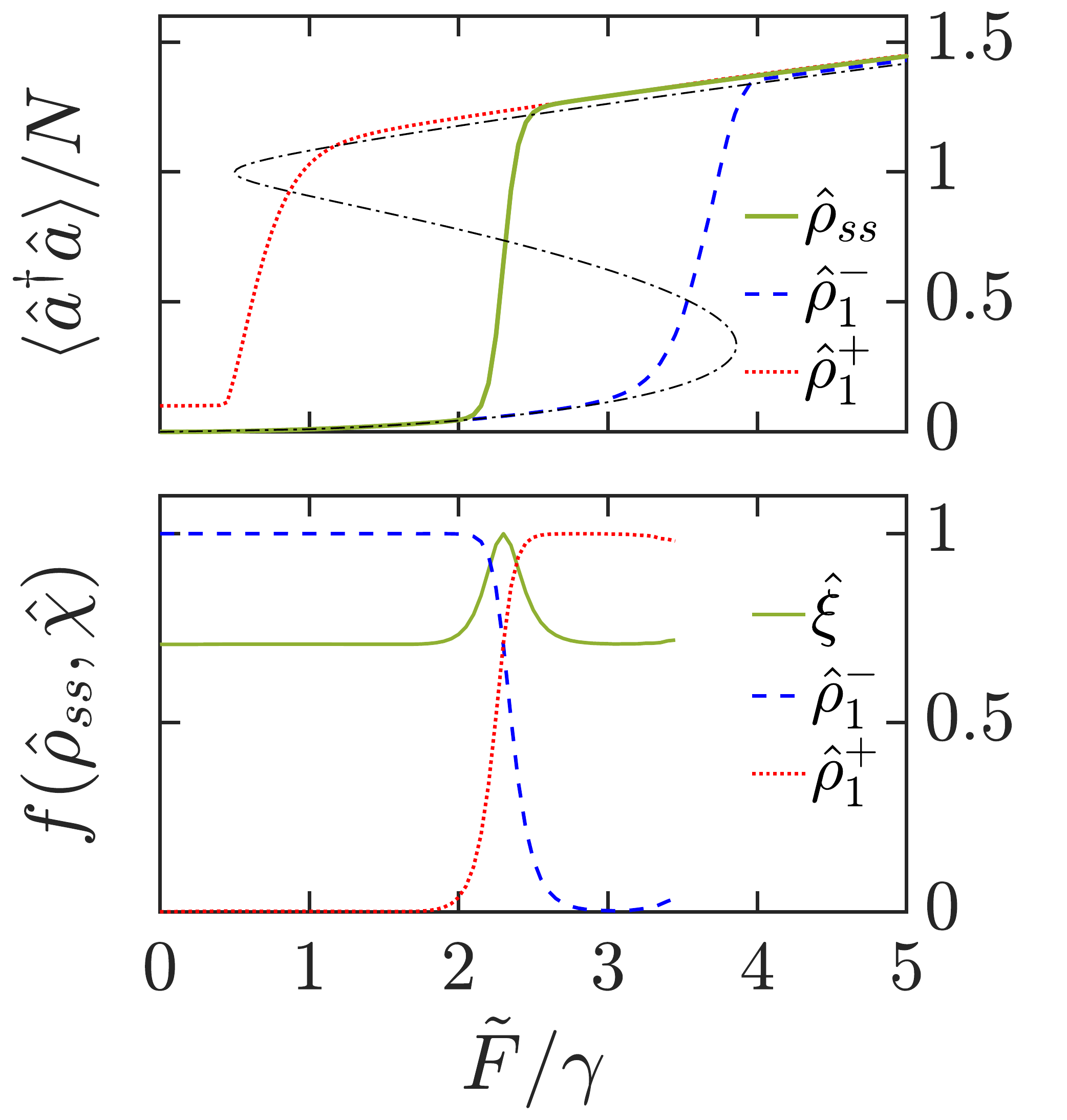} 
	\caption{Top panel: Average number of photons according to $\sss$ and   $\hat{\rho}^\pm$ as a function of the rescaled driving $\tilde{F}/\gamma$ for $N=10$ (see definitions in the main text).
		The dotted line indicates the Gross-Pitaevskji prediction.
		Bottom panel: Fidelity $f$ between the steady-state density matrix $\sss$ and a density matrix $\hat{\chi}=\hat{\rho}^+$, $\hat{\rho} ^-$, $\hat{\xi}$ [for $\hat{\xi}=(\hat{\rho}^-+ \hat{\rho}^+)/2$] as a function of the rescaled driving $\tilde{F}/\gamma$. The transition can be seen as a switching from a region where $\hat{\rho} ^-$ describes the system to one where the physics is dominated by $\hat{\rho}^+$.
		Even if the region of phase coexistence in $\sss$ is very narrow, $\eig{1}$ describes the physics in a larger region.
		Parameters are set as in Fig. \ref{fig:Kerr_resonator_transition}.
	}
	\label{fig:bistability}
\end{figure}

The first example which we discuss is the general model of a single driven-dissipative Kerr nonlinear resonator, for which an exact solution of $\sss$ exists \cite{DrummondJPA80}.
In a reference frame rotating at the coherent pump frequency $\omega_p$, the Hamiltonian of this system is
\begin{equation}\label{Eq:HCompleteOur}
\hat{H}=-\Delta \hat{a}^\dagger \hat{a}+\frac{U}{2} \hat{a}^\dagger \hat{a}^\dagger \hat{a} \hat{a}
+F (\hat{a}^\dagger + \hat{a}),
\end{equation}
where $\Delta=\omega_p-\omega_c$ is the pump-cavity detuning, $F$ is the driving amplitude and $U$ quantifies the Kerr nonlinearity.
The operators $\hat{a}^\dagger$ and $\hat{a}$ are the bosonic creation and annihilation operators, respectively.
The corresponding Lindblad master equation reads
\begin{equation}\label{Eq:MasterEquationOur}
\partial_t \hat{\rho}(t)=
- \ii \left[\hat{H},\hat{\rho}(t)\right]+ \frac{\gamma}{2} \mathcal{D}[\hat{a}] \hat{\rho}(t),
\end{equation}
where $\gamma$ is the dissipation rate of the cavity mode.
The properties of this model, and the emergence of a first-order phase transition, have been extensively discussed in Refs. \cite{BartoloPRA16,CasteelsPRA16,CasteelsPRA17-2}.
A well-defined thermodynamic limit is obtained for $\vert F \vert \to + \infty$ while keeping $U \vert F \vert ^2$ constant \cite{CasteelsPRA17-2}. 
This is equivalent to expressing nonlinearity and driving amplitude in the following form:
\begin{equation}
U = \tilde{U}/N,  \qquad F = \tilde{F} \sqrt{N},
\end{equation}
and letting $N \to + \infty$. 
In Fig.~\ref{fig:Kerr_resonator_transition} we study numerically the emergence of the first-order phase transition by increasing $N$.
The top panel shows the mean value of $\braket{\hat{a}^\dagger \hat{a}}/N = \Tr{\sss \hat{a}^\dagger \hat{a}}/N$ as a function of  $\tilde{F}/\gamma$. 
The middle panel shows the rescaled Liouvillian gap $-\Re{\lambda_1/\gamma}$ as a function of  the rescaled driving amplitude.
Such Liouvillian gap tends to zero in the thermodynamic limit $N \to + \infty$, while $\Im{\lambda_1}=0$ around the critical point also for finite $N$.
The bottom panel of Fig.~\ref{fig:Kerr_resonator_transition} presents a study of the fidelity between the steady state $\sss$ and the matrix $\hat{\xi}= (\eig{1}^+ + \eig{1}^-)/2$, obtained by the spectral decomposition of $\eig{1}$ [Eq.~\eqref{Eq:DecoRhoReal}].
We recall that the fidelity is defined as $f(\hat{\rho}, \hat{\xi})=\Tr{\sqrt{\sqrt{\hat{\rho}} \; \hat{\xi} \sqrt{\hat{\rho}}}}$. 
A fidelity equal to $1$ indicates that the two states are identical. 
As the thermodynamical parameter $N$ increases, we notice two important effects: (i) in the region in which the Liouvillian gap is minimal the fidelity is maximal; (ii) the region in which $\sss$ and $\hat{\xi}$ are close becomes narrower and narrower. This is consistent with our general results which are exact in the thermodynamic limit.

It is interesting now to connect our findings with the results predicted by mean-field theories.
A Gross-Pitaevskii-like mean-field approximation for the driven-dissipative Kerr model is known to exhibit bistability, while the full quantum solution is always unique \cite{DrummondJPA80}.
In the same way, a Gutzwiller-mean-field theory predicts multiple solutions \cite{LeBoitePRL13, BiondiPRA2017}.
In Fig.~\ref{fig:bistability}, we investigate the properties of the exact steady state $\sss$ and of the density matrices $\eig{1}^+$ and $\eig{1}^-$ for a system with $N=10$ as a function of the rescaled driving amplitude $\tilde{F}/\gamma$.
In the top panel, we plot the mean photon density $\braket{\hat{a}^\dagger \hat{a}}/N=\Tr{\hat{a}^\dagger \hat{a} \, \hat{\chi}}/N$, for $\hat{\chi}=\sss, \,\eig{1}^+, \, \eig{1}^-$ as indicated in the legend.
To further characterize the nature of $\eig{1}^\pm$ in the phase transition, in the bottom panel we plot the fidelity between $\sss$ and $\hat{\chi}=\hat{\xi}, \, \eig{1}^+, \, \eig{1}^-$ (where $\hat{\xi}=(\eig{1}^++\eig{1}^-)/2$).
For $\tilde{F}< \tilde{F}_c$, $\sss$ is almost exactly $\eig{1}^-$. 
Around the critical point $F\simeq F_c$, $\sss$ becomes an equal mixture of $\eig{1}^+$ and $\eig{1}^-$. The maximal mixed character occurs for $\tilde{F}= \tilde{F}_c$.
Finally, for $\tilde{F}>\tilde{F}_c$, the density matrix becomes very close to $\eig{1}^+$. 
This analysis allows us to interpret the two stable solutions predicted by the mean-field approach in terms of the metastable states which compose $\eig{1}$.

\subsection{The driven-dissipative resonator with two-photon pumping}
	
	As an example of second-order dissipative phase transition with symmetry breaking,  we will consider the driven-dissipative Kerr model with two-photon pumping and losses. In a reference frame rotating at the parametric pump frequency, the Hamiltonian of this system is \cite{BartoloPRA16,BartoloEPJST17}
	\begin{equation}\label{Eq:H2CompleteOur}
	\hat{H}=-\Delta \hat{a}^\dagger \hat{a}+\frac{U}{2} \hat{a}^\dagger \hat{a}^\dagger \hat{a} \hat{a}
	+\frac{G}{2} (\hat{a}^\dagger \hat{a}^\dagger + \hat{a} \hat{a}),
	\end{equation}
	where $G$ is the two-photon driving amplitude.
	This time, in addition to the Hamiltonian and to the one-photon dissipation superoperator $\mathcal{D}\left[\hat{a}\right]$, we will consider also a two-photon dissipation channel with rate $\eta$.  The corresponding Lindblad master equation reads:
	\begin{equation}\label{Eq:MasterEquation2}
	\partial_t \rho(t)=
	- \ii \left[\hat{H},\hat{\rho}(t)\right]+ \frac{\gamma}{2} \mathcal{D}[\hat{a}] \hat{\rho}(t) + \frac{\eta}{2} \mathcal{D}[\hat{a}^2] \hat{\rho}(t).
	\end{equation}
	The analytical solution of the steady state of this model has been provided in Ref.~\cite{MingantiSciRep16}, and the emergence of first- and second- order phase transitions (according to the value of $\Delta$) has been discussed in Ref.~\cite{BartoloPRA16}.
	The emergence of a similar symmetry breaking has been also observed in an equivalent classical system \cite{LeuchPRL16}.
	The thermodynamic limit of this model is obtained 
	by expressing $U$ and $\eta$ as 
	\begin{equation}
	U = \tilde{U}/N, \qquad \eta = \tilde{\eta} /N,
	\end{equation}
	and considering the limit $N \to + \infty$. In this way the ratio $U/\eta$ is kept constant.
	This model has a discrete $Z_2$ symmetry, resulting from the invariance under the transformation $\hat{a}\rightarrow -\hat{a}$.
	The corresponding superoperator $\mathcal{Z}_2$ is:
	\begin{equation}
	\mathcal{Z}_2= e^{\ii \pi \hat{a}^\dagger \hat{a}} \bigcdot e^{-\ii \pi \hat{a}^\dagger \hat{a}},
	\end{equation}
	with $\mathcal{Z}_2 \sss=\sss$.
	
\begin{figure}[t!]
		\centering
		\includegraphics[width=0.98\linewidth]{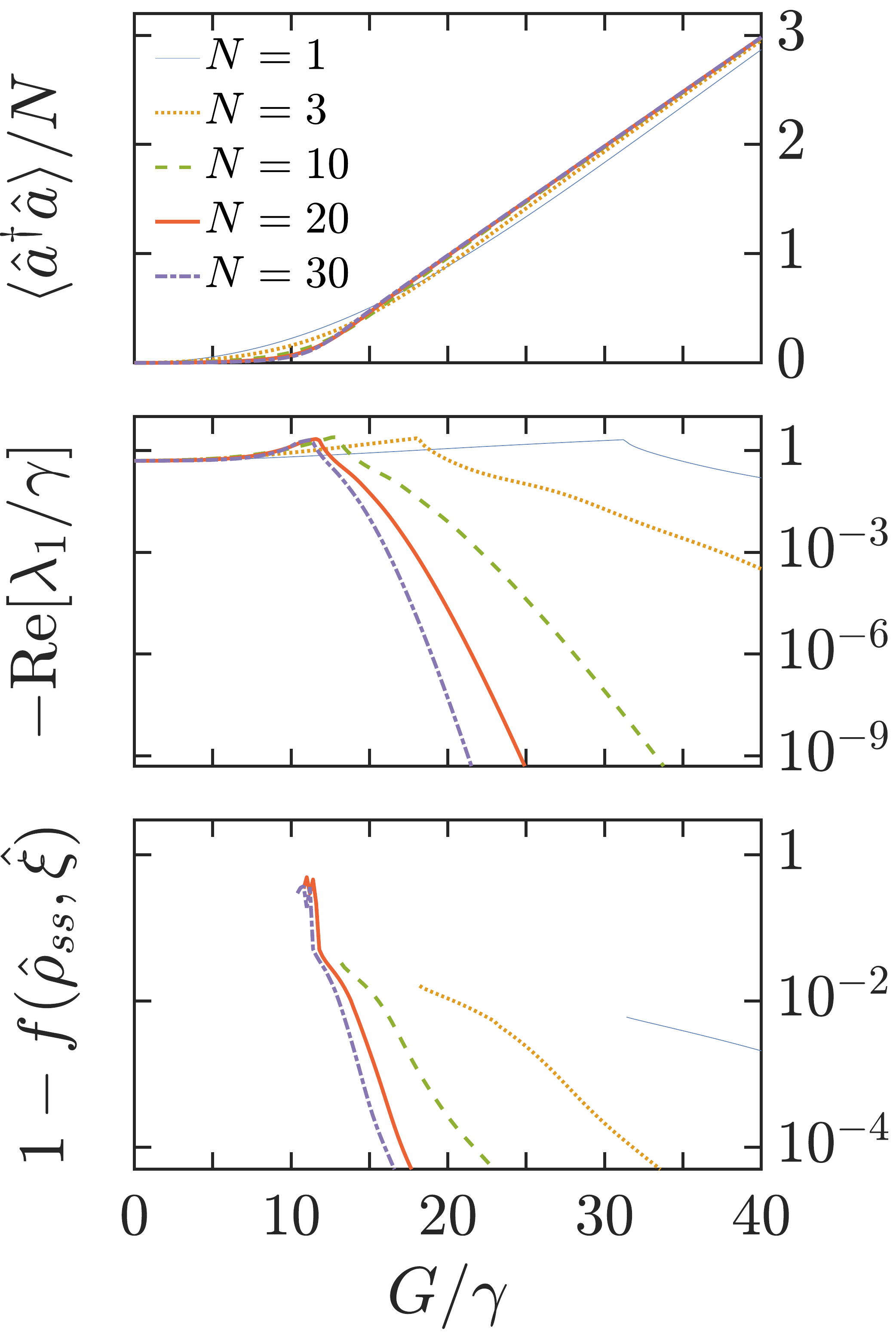}
\caption{
Numerical results for the driven-dissipative two-photon Kerr  model.
Top panel: Rescaled number of photons $\braket{\hat{a}^\dagger \hat{a}}/N $ as a function of the rescaled driving $G/\gamma$ for different values of $N$.
Middle panel: -${\rm Re} [\lambda_1/\gamma]$ (Liouvillian gap) for different values of $N$.
Bottom panel: The error $1-f$, where $f$ is the fidelity between the steady state density matrix $\sss$ and $\hat{\xi} = (\eig{1}^+ + \eig{1} ^-)/2$.
The curves are shown in the region where $\lambda_1$ is purely real.
Parameters: $\Delta/\gamma=-10$, $\tilde{U}/\gamma=10$, $\tilde{\eta}/\gamma=1.0$.}
		\label{fig:Kerr_2ph_resonator_transition}
	\end{figure}

In Fig.~\ref{fig:Kerr_2ph_resonator_transition} we show the emergence of a second-order phase transition by increasing the value of $N$.
The top panel shows $\braket{\hat{a}^\dagger \hat{a}}/N = \Tr{\sss \hat{a}^\dagger \hat{a}}/N$ as a function of $G/\gamma$.
In the middle panel we show the rescaled Liouvillian gap $-\Re{\lambda_1/\gamma}$ as a function of the rescaled pump amplitude.
The abrupt change in the behavior of $\lambda$ indicates the onset of the phase transition.
In the whole region of broken symmetry, the gap is much smaller than $\gamma$ and $\lambda_1$ is real,
while $\eig{1}$ is a traceless Hermitian matrix which belongs to the odd symmetry sector of $\mathcal{Z}_2$   ($\mathcal{Z}_2\eig{1}=-\eig{1}$). 
The states $\eig{1}^+$ and $\eig{1}^-$ obtained via the spectral decomposition of $\eig{1}$ are such that $\mathcal{Z}_2 \eig{1}^+ = \eig{1}^-$.
As it has been shown in Sec. \ref{Sec:Secon_order}, in the symmetry-broken region, $\sss$ can be constructed as a symmetric mixture of $\eig{1}^+$ and $\eig{1}^-$. As shown in the bottom panel of Fig. \ref{fig:Kerr_2ph_resonator_transition}, this gives an excellent approximation for the finite-sized systems considered here. Remarkably, this expression for $\sss$ remains very accurate even quite far from the thermodynamic limit.

	\begin{figure}[h!]
		\centering
		\includegraphics[width=0.98\linewidth]{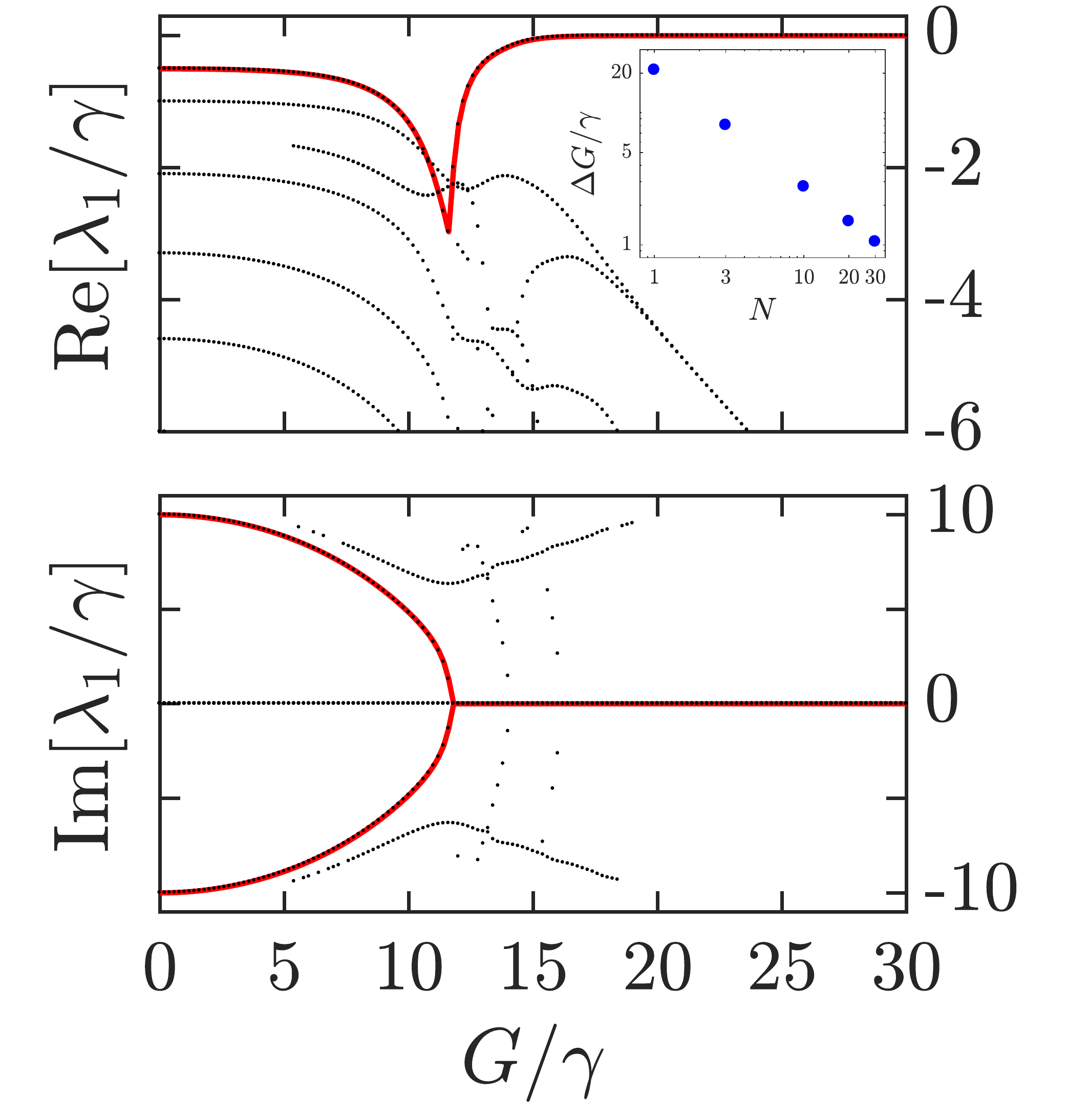}
		\caption{
			Liouvillian spectrum in the two-photon Kerr model for $N=20$. 
			Top and bottom panels: Real and imaginary part of the eigenvalues of $\supmat{\L}$. 
			The dots represent the 10 smallest-modulus eigenvalues obtained by numerical diagonalization.
			The red lines are a guide for the eye indicating the two eigenvalues which merge into $\lambda_1$ for $G\gg \gamma$.
			Inset: A log-log plot of $\Delta G= (G_B(N) - G_c)$ (defined in the text) as a function of the parameter $N$, showing the power-law behavior $\Delta G = A \, N^{-\eta}$ with $A=21.1 \pm 0.2$ and $\eta=0.881 \pm 0.006$.
	        In the thermodynamic limit, the bifurcation point $G_B(N)$ and the critical point $G_c$ coincide.
        Same parameters as in Fig.~\ref{fig:Kerr_2ph_resonator_transition}.}
		\label{fig:Kerr_2ph_resonator_exceptional}
	\end{figure}

\begin{figure}[h!]
	\centering
	\includegraphics[width=0.98\linewidth]{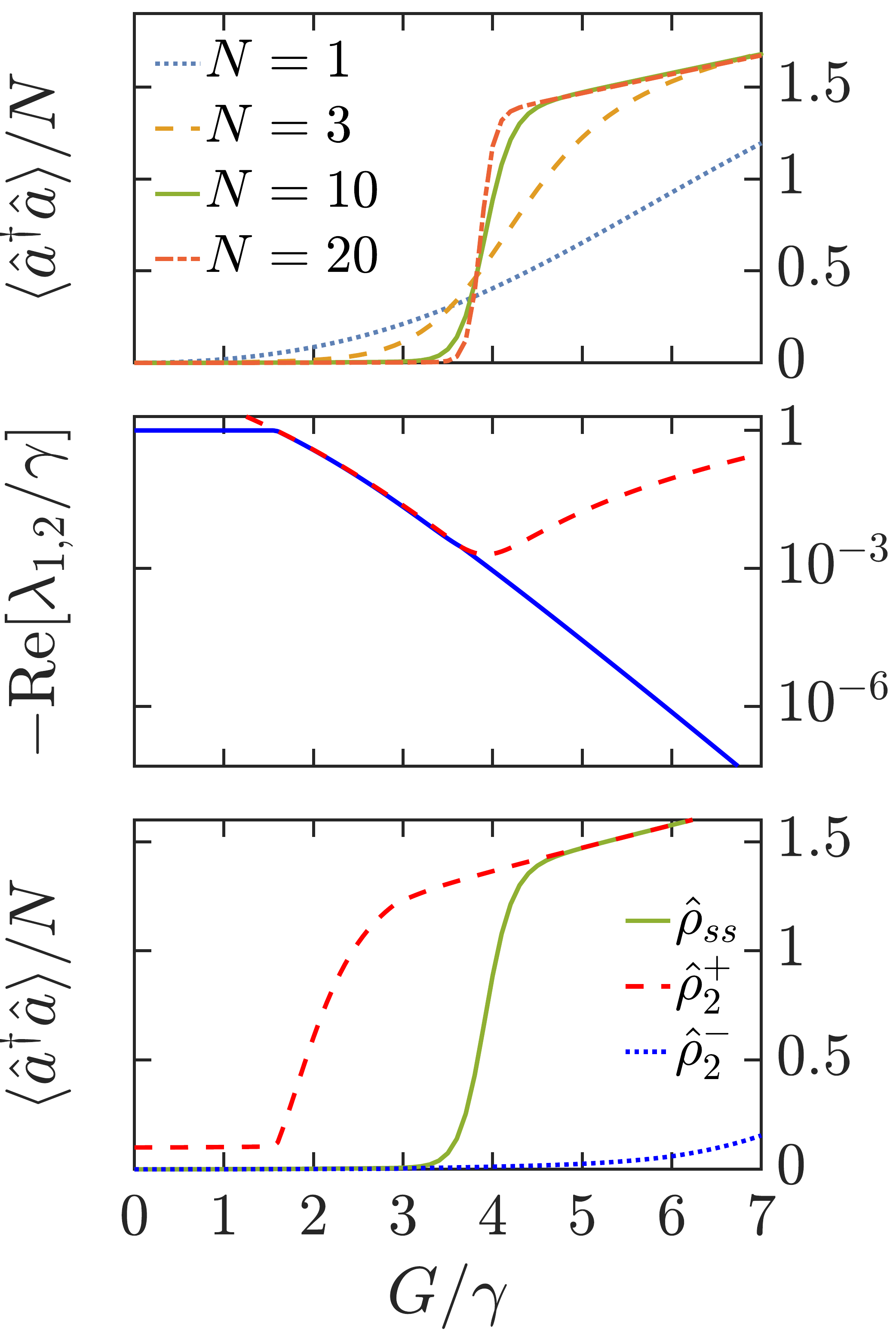}
	
	\caption{Study of the first-order phase transition with symmetry breaking in the two-photon Kerr resonator.
		Top panel: Rescaled number of photons $\braket{\hat{a}^\dagger \hat{a}}/N$ as a function of the rescaled driving $G/\gamma$ for different values of $N$.
		Middle panel: Real part of $\lambda_{1,2}$ rescaled by $\gamma$ for $N=10$.
		The two branches of Liouvillian eigenvalues lead to the first order phase transition (red) and a symmetry breaking (blue).
		Bottom panel: Average number of photons according to $\sss$ and $\eig{2}^\pm$.
		Parameters: $\Delta/\gamma=10$, $U/\gamma=10$, $\eta/\gamma=1.0$.}
	\label{fig:Kerr_2ph_first_order_transition}
\end{figure}

In order to characterize the abrupt change in the behavior of $\lambda_1$, which becomes discontinuous for $N\to +\infty$, we plot part of the full spectrum of $\supmat{\L}$ for $N=20$ across the critical point.
In the top panel of Fig.~\ref{fig:Kerr_2ph_resonator_exceptional}, we show the real part of the spectrum, while the bottom one reports the imaginary part. Starting from the imaginary part, we clearly see that there is a point in which two complex-conjugate eigenvalues (highlighted by the red line) become real. We call $G_B(N)$ the point at which this bifurcation happens.
Looking at the top panel, this merging is associated with a change in the behavior of the real part of those eigenvalues, which split and bifurcate.
The one approaching zero is responsible for the phase transition and its associated eigenvector becomes $\eig{1}=\eig{1}^+ - \eig{1}^-$ for $G>G_B(N)$.
As we saw in Fig~\ref{fig:Kerr_2ph_resonator_transition}, it is not clear where the gap starts to close, but one might guess that it happens when the two eigenvalues bifurcate. To test this conjecture, in the inset we plot, as a function of $N$, the scaling of the bifurcation point $\Delta G= G_B(N)-G_c$, where $G_c$ is the critical point extrapolated via the study of the analytic solution for $N=1000$.
Indeed, the clear power-law decay of this quantity demonstrates that the onset of this transition can be understood in terms of a merging of two eigenvalues. 
The emergence of criticality is thus to be associated with a {\it touching} of two eigenvalues in the complex plane.
This fact, together with the emergence of a discontinuity in $\lambda_1$ for $N\to\infty$, implies that, at the bifurcation point, the Liouvillian becomes non diagonalizable, resulting in a Jordan structure. This leads to a non-exponential relaxation dynamics at criticality.
To better understand this behavior, in App.~\ref{app:jordan} we study an exactly-solvable two-level system which admits a Jordan-block structure for a specific choice of parameters.
	
Up to now, we considered the case in which an eigenvalue of the symmetry sector $L_{-1}$ approaches zero, which gives rise to a symmetry breaking without inducing first-order discontinuities in $\sss$.
The two-photons Kerr model is known to present also a first-order phase transition with symmetry breaking for $\Delta > 0$ \cite{BartoloPRA16}.
Indeed, together with the emerging of a zero in $L_{-1}$, the symmetry sector $L_{1}$ acquires two zero eigenvalues of $\L$: one associated with $\sss$, the other with an eigenmatrix whose eigenvalue touches zero only at the critical point.
This allows a discontinuous behavior of $\sss$ with symmetry breaking.
In Fig.~\ref{fig:Kerr_2ph_first_order_transition} we plot the behavior of the system in such regime.
The top panel shows the emergence of a first-order phase transition in the rescaled density.
In the middle panel, we plot the real part of the two eigenvalues of the Liouvillian with the smallest modulus.
One presents the phenomenology we expect from a symmetry breaking: $-\Re{\lambda_1}\ll\gamma$ in the symmetry-broken phase $G\geq G_c(N)$. The other is responsible for the discontinuous first-order behavior: $-\Re{\lambda_2}\ll\gamma$ only for $G\simeq G_c(N)$.
Indeed, we tested that $\eig{1}$ (associated with $\lambda_1$) satisfies $\mathcal{Z}_2 \eig{1}=-\eig{1}$.
Moreover, $\mathcal{Z}_2 \eig{1}^\pm = \eig{1}^\mp$ and $\sss \simeq (\eig{1}^+ + \eig{1}^-)/2$.
As for $\eig{2}$, $\mathcal{Z}_2\eig{r}=\eig{r}$ and it cannot be associated with a symmetry breaking.
In the bottom panel we test the structure of $\sss$ in connection to the spectral decomposition of $\eig{2}$: the first-order phase transition can be interpreted as a switch between $\eig{2}^-$ and $\eig{2}^+$.
The symmetry breaking emerges in the fact that $\eig{2}^+\simeq\left(\eig{1}^+ + \eig{1}^-\right)/2$.
In conclusion, in these specific numerical examples we recover all the features predicted by our general theory.

	
\section{Conclusions}
\label{Sec:Conclusion}

In this article, we have presented theoretical results for first- and second-order dissipative phase transitions.
Within a general formalism, we have determined the structure of the density matrix in the vicinity of a critical point. In particular, due to the closure of the Liouvillan gap at the critical point, we have shown how the the steady-state density matrix is directly related to the eigenmatrix of the Liouvillian superoperator corresponding to the eigenvalue $\lambda_1$ (the one with the smallest absolute value of the real part). We have illustrated our general results by considering two specific quantum optical models, where the emergence of a dissipative phase transition can be studied analytically and numerically. 
Our work provides a general insight into dissipative phase transitions.
Moreover, it gives precise constraints for variational methods \cite{WeimerPRL2015,BanulsPRL15} to describe critical phenomena in open quantum systems, whose corresponding ansatz matrices must satisfy the relations derived in this work.

\acknowledgements
We acknowledge discussions with V. Albert, G. Orso, D. Rossini, R. Rota, and M. Vogel.
We acknowledge support from ERC (via Consolidator Grant CORPHO No. 616233) and ANR (via the grant UNIQ ANR-16-CE24-0029).

\appendix


\section{Proofs}
\label{App:Demonstrations}

\subsection{Remarks on the Liouvillian diagonalisability}
\label{App:DiagonalisableLiouvillian}

Generally, the Liouvillian is a non-Hermitian superoperator with a holomorphic dependence on the system parameter(s) $\zeta$.
Therefore, there might exist values of $\zeta$ for which $\L(\zeta)$ is not diagonalizable: this implies the existence of a degenerate eigensubspace.

The eigenvalues ${\lambda_i(\zeta)}$ of $\L(\zeta)$ can be obtained via the resolution of the characteristic equation $\det \left(\Lmat(\zeta) - \lambda_i(\zeta) \mathbb{I} \right) =0$.
A well-known result of function theory \cite{KatoBOOK} guarantees that the roots of this equation are branches of analytic functions of $\zeta$ with, at most, algebraic singularities.
Therefore, the number $s$ of distinct eigenvalues of $\L(\zeta)$ is a constant except in a countable number of points.
This ensures that if the Liouvillian has a simple spectrum on a finite region of the parameter space, it will be diagonalizable for any $\zeta$, except the countable exceptional points.
For all the systems considered in this work, this condition is fulfilled far from the thermodynamic limit.

\subsection{Proofs of the Lemmas in \ref{subsec:eigenvalues}}
\label{App:ProvesOnLiouvillian}

\textbf{Lemma 3:} If $\L\eig{i}=\lambda_i\eig{i}$ then  $\L\eig{i}^\dagger=\lambda_i^*\eig{i}^\dagger$. This implies that, if $\eig{i}$ is Hermitian then $\lambda_i$ has to be real.
Conversely, if $\lambda_i$ is real and of degeneracy 1, $\eig{i}$ is Hermitian.
If $\lambda_i$ has geometric multiplicity $n$ and $\L$ is diagonalizable, it is always possible to construct $n$ Hermitian eigenmatrices of $\L$ with eigenvalue $\lambda_i$.
\newline
\textbf{Proof:}
Thanks to the master equation we have:
\begin{equation}
\label{eq:hermitian_is_eigenmatrix}
\begin{split}
\mathcal{L}\eig{i}^\dagger&=-\ii \left[\hat{H}, \eig{i}^\dagger\right]+\frac{\gamma}{2}\left(2 \hat{a}\eig{i}^\dagger \hat{a}^\dagger -\hat{a}^\dagger \hat{a} \eig{i}^\dagger - \eig{i}^\dagger \hat{a}^\dagger \hat{a} \right)\\
&=\left(-\ii \left[\hat{H}, \eig{i}\right]+\frac{\gamma}{2}\left(2 \hat{a}\eig{i}\hat{a}^\dagger -\hat{a}^\dagger \hat{a} \eig{i}- \eig{i} \hat{a}^\dagger \hat{a} \right)\right)^\dagger\\ 
&=\left(\mathcal{L}\eig{i}\right)^\dagger=\lambda_i^*\eig{i}^\dagger.
\end{split}
\end{equation}
If $\eig{i}$ is Hermitian, we have $\lambda_i\eig{i}=\mathcal{L}\eig{i}=\mathcal{L}\eig{i}^\dagger=\lambda_i^*\eig{i}^\dagger=\lambda_i^*\eig{i}$. Thus, we can conclude $\lambda_i=\lambda_i^*$.
Conversely, in the case in which $\lambda_i\in \mathbb{R}$ is a simple eigenvalue (i.e. with degeneracy 1), we can conclude that $\eig{i}=\eig{i} ^\dagger$, and thus $\eig{i}$ is Hermitian.
If the eigenvalues have geometric multiplicity $n$, it may happen that for some eigenmatrices $\eig{i} ^\dagger \neq \eig{i}$.
From Eq.~\eqref{eq:hermitian_is_eigenmatrix} it follows $\L \eig{i}^\dagger= \lambda_i \eig{i}^\dagger$.
In this case, we can consider the matrices $\left(\eig{i}+\eig{i}^\dagger\right)/2$ and $\ii  \left(\eig{i}-\eig{i}^\dagger\right)/2$, which are Hermitian by construction, and whose eigenvalue is $\lambda_i$.

\textbf{Lemma 4:} If $\lambda_i=0$ has degeneracy $n$, then there exist $n$ independent eigenvectors of the Liouvillian (the algebraic multiplicity is identical to the geometrical one). Therefore, there exist $n$ different steady states towards which the system can evolve, depending on the initial condition.
\newline
\textbf{Proof:}
We will prove this lemma by contradiction. Let us suppose that the algebraic multiplicity is greater than the geometrical one (see \cite{Note2}).
Since the dimension of the reduced space is $n$, we can write the Liouvillian as a matrix acting on a basis of vector in this reduced space, i.e. the invariant space of $\lambda_0$ has a finite dimension.
Since we can write the Liouvillian as a matrix, this means that we can put in its canonical Jordan form.
In other words the Liouvillian acting on the vectors of this subspace can be decomposed in a diagonal part $\supmat{\Lambda}_0$ and a nilpotent matrix $\supmat{N}$ via a similarity transformation $S$:
\begin{equation}
\supmat{\L}_{\lambda_0}=S^{-1} \left(\supmat{\Lambda}_0 + \supmat{N}\right) S= S^{-1} \left[\begin{array}{c c c c c  c}
\lambda_0 & 1 & 0& \cdots  & 0 & 0 \\
0 &\lambda_0 &1& \cdots & 0 & 0 \\
\vdots & \ddots & \ddots & \ddots & \vdots & \vdots \\
0& 0& 0 & \cdots  & \lambda_0 &  1 \\
1& 0  & 0 & \cdots& 0  & \lambda_0\\
\end{array}
\right] S.
\end{equation}
Of course, the new basis of vectors obtained by the nonunitary transformation $S$  may not be orthonormal.
The time evolution of the system is given by $e^{\Lmat t}$, and since $\supmat{\Lambda}_0$ and $\supmat{N}$ commute, one has 
\begin{equation}
\begin{split}
e^{\supmat{\L}_{\lambda_0} t} & =S^{-1}  e^{\supmat{\Lambda}_0 t} e^{\supmat{N}t} S \\
&= S^{-1} e^{\lambda_0 t} (\mathds{1}+\supmat{N}t+\frac{(\supmat{N}t)^2}{2}+ \dots \frac{(\supmat{N}t)^n}{n!})  S\\ 
& = S^{-1} e^{\lambda_0 t}\left[\begin{array}{c c c c c  c} 
1 & t & \frac{t^2}{2}& \cdots  & \frac{t^{n-1}}{(n-1)!} & \frac{t^n}{n!} \\
0 & 1 & t & \cdots & \frac{t^{n-2}}{(n-2)!} & \frac{t^{n-1}}{(n-1)!} \\
\vdots & \ddots & \ddots & \ddots & \ddots & \ddots \\
0& 0& 0 & \cdots  & 1 &  t \\
1& 0  & 0 & \cdots& 0  & 1\\
\end{array}
\right] S.
\end{split}
\end{equation}
Since $\lambda_0 = 0$, the previous expression clearly will cause the dynamics to diverge, proving the absurd.
We stress that this reasoning cannot be directly extended to $\lambda_i\neq0$ nor to infinite degeneracies $n\to\infty$.

\subsection{Vanishing of $\lambda_{1}$ associated to a first-order phase transition}
\label{app:first_order_gap_demonstration}

In Sec.~\ref{Sec:First_order_Phase_transition} we proved that if there is a jump in one observable at the critical point $\zeta=\zeta_c$, than $\lambda_{1}=0$.
Here, we prove that the last condition is also sufficient; i.e., $\lim\limits_{\zeta\to \zeta_c} \lambda_1(\zeta) = 0$ implies a first-order phase transition.

We will prove this statement by contradiction. Let us suppose that even if $\lambda_1=0$ there is no phase transition.
From the definition~\eqref{Eq:DPTDefinition}, we deduce that for any operator $\hat{o}$ in $\mathcal{H} \otimes \mathcal{H}$, $\braket{\hat{o}(\zeta)}$ is continuous in $\zeta_c$.
Hence, we have that also $\eig{0}(\zeta)$ is continuous.
From Lemmas 3 and 4 of Sec.~\ref{subsec:eigenvalues}, the eigenstate $\eig{1}(\zeta_c)$, being associated with $\lambda_1(\zeta_c)=0$, exists and is Hermitian.
By exploiting its spectral decomposition, we can write $\eig{1}(\zeta_c) = (\eig{1}^+(\zeta_c) - \eig{1}^- (\zeta_c))/\sqrt{2}$ (we stress that here we have  $\|\eig{1}^\pm(\zeta_c) \|=1$, and $\braket{\eig{1}^+(\zeta_c),\eig{1}^-(\zeta_c)}=0$ by construction).

The first part of the proof is to show that, $\eig{0}(\zeta_c) = (\eig{1}^+(\zeta_c) + \eig{1}^-(\zeta_c) )/\sqrt{2}$.
Indeed, $\|\eig{1}(\zeta_c)\|=1$ and $e^{\L t} \eig{1} =\eig{1} $.
Thus, exploiting the triangular inequality, we have:
\begin{equation}
\begin{split}
1= \|\eig{1}(\zeta_c)\|^2 &= \left\| e^{\L t} \eig{1}(\zeta_c) \right\|^2 = \left\| e^{\L t} \frac{\eig{1}^+(\zeta_c) - \eig{1}^- (\zeta_c)}{\sqrt{2}} \right\| ^2  
\\ &\leq \frac{\left\| e^{\L t} \eig{1}^+(\zeta_c)\right\|^2  + \left\| e^{\L t} \eig{1}^-(\zeta_c) \right\|^2 }{2} \leq 1.
\end{split}
\end{equation}
It follows that $\left\| e^{\L t} \eig{1}^\pm (\zeta_c)\right\| = 1$ for every time $t$.
Hence, $\eig{1}^\pm(\zeta_c)$ must be a linear superposition of eigematrices of the Liouvillian with zero eigenvalue. 
Considering that $\eig{1}(\zeta_c) = (\eig{1}^+(\zeta_c) - \eig{1}^-(\zeta_c))/\sqrt{2}$, $\| \eig{0}(\zeta_c)\| =1$, and $\braket{\eig{1}^+(\zeta_c),\eig{1}^-(\zeta_c)}=0$, we obtain $\eig{0}(\zeta_c)=(\eig{1}^+(\zeta_c) + \eig{1}^-(\zeta_c))/\sqrt{2}$.

Having proved the first part, let us consider the eigendecomposition of $\eig{1}(\zeta)$ around $\zeta_c$.
Except at the critical point, we have $\lim_{t \to \infty}e^{\L t} \eig{1}^\pm(\zeta)=\eig{0}(\zeta)/\Tr{\eig{1}^\pm}$.
But, by hypothesis, all function are continuous, hence:
\begin{equation}
\begin{split}
\eig{1}^\pm (\zeta_c) & =\lim_{\zeta\rightarrow \zeta_c}\lim_{t \to \infty}e^{\L t} \eig{1}^\pm(\zeta)=\lim_{\zeta\rightarrow \zeta_c}\eig{0}(\zeta)/\Tr{\eig{1}^\pm} 
\\ &= \frac{\eig{1}^+(\zeta_c) + \eig{1}^-(\zeta_c)}{2}.
\end{split}
\end{equation}
Consequently, we find that at the critical point, $\eig{1}^+(\zeta_c)= \eig{1}^-(\zeta_c)$.
This statement would require that at $\zeta=\zeta_c$ $\eig{1}(\zeta_c)=0$.
This statement is absurd, since Lemma 4 of Sec.~\ref{subsec:eigenvalues} guarantees that $\eig{1}(\zeta_c)$ is a well-defined eigenvector of the Liouvillian.
Therefore, by the absurd, we deduce that the function $\eig{0}(\zeta)$ can not be continuous at $\zeta=\zeta_c$.


\section{Nonexponential decays and Jordan blocks}
\label{app:jordan}

In this appendix, we provide a simple analytic example of a non-exponential decay associated with a Jordan block structure of the Liouvillian.
Let us consider a spin-$1/2$ subjected to the action of two competing decay channels whose evolution obeys ($\hbar=1$) \cite{SarandyPRA05}
\begin{equation}
\label{me}
\partial_t \hat{\rho}(t) = \mathcal{L}\hat{\rho}(t)= -i [\hat{H}, \hat{\rho}(t)]+\frac{\epsilon}{2}\mathcal{D}[\hat\sigma^-]\hat{\rho}(t)+\frac{\gamma}{2}\mathcal{D}[\hat\sigma^x]\hat{\rho}(t),
\end{equation}
where $\hat H=\frac{\omega}{2}\hat{\sigma}^z$.
The steady-state density matrix can be obtained as
\begin{equation}
\sss=\frac{1}{2\gamma+\epsilon}\begin{pmatrix} 
	\gamma & 0 \\
	0 & \gamma+\epsilon 
\end{pmatrix}. 
\end{equation}
The eigenvalues of $\L$ are
\begin{equation}
\begin{split}
\lambda_0 =& 0, \\
\lambda_1 =& -\gamma -\frac{\epsilon }{2} + \sqrt{\gamma ^2-\omega ^2}, \\
\lambda_2 =& -\gamma -\frac{\epsilon }{2} - \sqrt{\gamma ^2-\omega ^2}, \\
\lambda_3=& -2 \gamma -\epsilon, \\
\end{split}
\end{equation}
which are associated with the following (unnormalized) eigenmatrices
\begin{widetext} 
\begin{equation}
	\eig{0}\propto\sss=\frac{1}{2\gamma+\epsilon}\begin{pmatrix} 
		\gamma & 0 \\
		0 & \gamma+\epsilon 
	\end{pmatrix}, \quad
	\eig{1}\propto\begin{pmatrix} 
		0 & \frac{\sqrt{\gamma ^2-\omega ^2}-i \omega }{\gamma } \\
		1 & 0 \\
	\end{pmatrix}, \quad
	\eig{2}\propto\begin{pmatrix}
		0 & -\frac{i \omega +\sqrt{\gamma ^2-\omega ^2}}{\gamma } \\
		1 & 0 \\
	\end{pmatrix}, \quad
	\eig{3}\propto\begin{pmatrix}
		-1 & 0 \\
		0 & 1 \\
	\end{pmatrix}.
\end{equation}
\end{widetext} 
The eigenmatrices $\eig{1,\, 2}$ describe the decay of the $\hat{\sigma}^{x,y}$ components with rate $\lambda_{1,\, 2}$, while  $\eig{3}$ is associated with $\hat{\sigma}^z$ and $\lambda_3$.

This simple model is particularly interesting since, according to the values of the couplings, it can display different relaxation dynamics toward the steady state:
\begin{itemize}
	\item If $\gamma>\omega$ the Liouvillian has $4$ real distinct eigenvalues (it is diagonalizable). 
		In this case, the decay at long times will be exponential. 
		The asymptotic decay rate is $\lambda_1$.
	
	\item If $\gamma<\omega$ the Liouvillian has $4$ distinct eigenvalues (it is diagonalizable), $2$ of which are complex conjugate ($\lambda_1=\lambda_2^*$). In this case, the decay at long times will be an exponential decay of magnitude ${\rm Re}[\lambda_1]={\rm Re}[\lambda_2]$ multiplied by an oscillating term given by ${\rm Im}[\lambda_1]$.
	
	\item If $\lambda=\omega$ we have $\lambda_1=\lambda_2$ and $\eig{1}=\eig{2}$: the Liouvillian is not diagonalizable but it can be written in a Jordan form.
\end{itemize}

The presence of a Jordan form has strong consequences for the long-time dynamics. 
Indeed, given a general initial state 
\begin{equation}
\label{ss}
\hat{\rho}(0)=\begin{pmatrix} 
	a & b \\
	b^* & 1-a 
\end{pmatrix},
\end{equation}
the decay of the observables $\hat\sigma^{x,\, y}$ is given by
\begin{equation}
\begin{split}
\Tr{\hat\sigma^x\hat{\rho}(t)} &= 2e^{-\lambda_1 t} (t \omega  ({\rm Re}[b]+{\rm Im}[b])+{\rm Re}[b]) \\
\Tr{\hat\sigma^y\hat{\rho}(t)}&=2e^{-\lambda_1 t} ( t \omega  ({\rm Re}[b]+{\rm Im}[b])- {\rm Re}[b]), 
\end{split}
\end{equation}
hence not exponential.
However, we stress that the asymptotic decay rate is $\lambda_3$ for $\hat\sigma^{z}$ (purely-exponential decay).
We also remark on the strong similarity between the behavior of $\lambda_{1,2}$ and that of the eigenvalues characterizing a second-order phase transition.
In both cases, a pair of two complex-conjugate eigenvalues becomes real in proximity to an exceptional point.


	\bibliography{bibliography}	
	
\end{document}